\documentclass[useAMS,usenatbib]{mn2e}
\usepackage{graphicx,fleqn}
\DeclareGraphicsExtensions{.eps,.ps,.eps.gz,.ps.gz,.eps.Z}
\title[Optical and Infrared Signatures of ULX's]{Optical and Infrared Signatures of Ultra-luminous X-ray Sources}

\author[Chris Copperwheat, Mark Cropper, Roberto Soria, Kinwah Wu]{Christopher Copperwheat$^{1}$, Mark Cropper$^{1}$ Roberto Soria$^{1,2}$ and Kinwah Wu$^{1}$\\
$^{1}$ Mullard Space Science Laboratory, University College London,
Holmbury St. Mary, Dorking, Surrey, RH5 6NT, UK\\
$^{2}$ Harvard-Smithsonian Center for Astrophysics, 60 Garden Street, Cambridge, MA 02138, USA
}

\date{Received: }

\begin{document}
 
\newcommand{\dg} {^{\circ}}
\outer\def\gtae {$\buildrel {\lower3pt\hbox{$>$}} \over
{\lower2pt\hbox{$\sim$}} $}
\outer\def\ltae {$\buildrel {\lower3pt\hbox{$<$}} \over
{\lower2pt\hbox{$\sim$}} $}
\newcommand{\ergscm} {ergs s$^{-1}$ cm$^{-2}$}
\newcommand{\ergss} {ergs s$^{-1}$}
\newcommand{\ergsd} {ergs s$^{-1}$ $d^{2}_{100}$}
\newcommand{\pcmsq} {cm$^{-2}$}
\newcommand{\ros} {{\it ROSAT}}
\newcommand{\xmm} {\mbox{{\it XMM-Newton}}}
\newcommand{\exo} {{\it EXOSAT}}
\newcommand{\sax} {{\it BeppoSAX}}
\newcommand{\chandra} {{\it Chandra}}
\newcommand{\hst} {{\it HST}}
\def\rchi{{${\chi}_{\nu}^{2}$}}
\newcommand{\Msun} {$M_{\odot}$}
\newcommand{\Mwd} {$M_{wd}$}
\newcommand{\Mbh}{$M_{\bullet}$}
\newcommand{\Lsun} {$L_{\odot}$}
\newcommand{\Rsun} {$R_{\odot}$}
\def\Mdot{\hbox{$\dot M$}}
\def\mdot{\hbox{$\dot m$}}
\def\mincir{\raise -2.truept\hbox{\rlap{\hbox{$\sim$}}\raise5.truept
\hbox{$<$}\ }}
\def\magcir{\raise -4.truept\hbox{\rlap{\hbox{$\sim$}}\raise5.truept
\hbox{$>$}\ }}

\maketitle

\begin{abstract}
We have constructed a model to describe the optical emission from ultra-luminous X-ray sources (ULXs). We assume a binary model with a black hole accreting matter from a Roche lobe filling companion star. We consider the effects of radiative transport and radiative equilibrium in the irradiated surfaces of both the star and a thin accretion disk. We have developed this model as a tool with which to positively identify the optical counterparts of ULXs, and subsequently derive parameters such as the black hole mass and the luminosity class and spectral type of the counterpart. We examine the dependence of the optical emission on these and other variables. We extend our model to examine the magnitude variation at infrared wavelengths, and we find that observations at these wavelengths may have more diagnostic power than in the optical. We apply our model to existing \hst \ observations of the candidates for the optical counterpart of ULX X-7 in NGC 4559. All candidates could be consistent with an irradiated star alone, but we find that a number of them are too faint to fit with an irradiated star and disk together. Were one of these the optical counterpart to X-7, it would display a significant temporal variation.
\end{abstract}


\maketitle

\section{INTRODUCTION}
\label{sec:intro}

The presence of point-like, non-nuclear and extremely luminous X-ray sources in local galaxies has been recognised for some time \citep{Fab03,Swartz04}. These ultra-luminous X-ray sources (ULX) can have luminosities $\sim 10^{40}$\ergss, and their true nature is yet to be understood. This luminosity greatly exceeds the Eddington luminosity $L_{Edd}$ of a 10 M$_{\odot}$ black hole (BH) if the emission is isotropic. This has led to the suggestion that the accreting object could be an intermediate mass black hole (IMBH) with mass $50-1000$ M$_{\odot}$ \citep{ColMus99,Maki00}. If so, these objects are a link between the established population of stellar mass black holes, and the supermassive black holes in active galactic nuclei. The presence of cool X-ray spectral components, and the timescales of rapid variability in some systems (e.g. NGC 4559 X-7, \citealt{Cr04}) is consistent with the IMBH scenario. However, the existence of IMBH is still under debate, as some observations appear contradictory. For example, some systems require high accretion disk temperatures in model fits to the X-ray data \citep{Ebisawa03}. It has also been noted that the emission is not likely to be isotropic, but collimated after to a greater or lesser extent \citep{King01, Kord02, Fabrika04}. Alternatively, it has also been argued by \citet{Bege02} that an accretion disk dominated by radiation pressure would exhibit strong density inhomogeneities on scales much smaller than the disk scale height, so that an inhomogeneous accretion disk could permit escaping flux to exceed $L_{Edd}$ by a factor of up to $\sim10-100$.

It seems increasingly likely that the ULX population is heterogeneous, with evidence to support beaming with stellar mass BHs in some cases and IMBH in others \citep{Fab04}. The QPO reported by \citet{Stroh03} is incompatible with the beaming hypothesis, for example, as is the association of some ULXs with diffuse H$\alpha$ nebulae, suggesting isotropic illumination of the interstellar medium by the ULX \citep{Pakull02,Miller03}. In this paper we assume an IMBH interpretation for ULX. Furthermore, we seek to describe the very brightest objects in the ULX population -- specifically, those that have luminosities of $10^{40}$\ergss \ or greater (e.g. in M82, \citealt{Matsu01,Kareet01}). We find that wind accretion from a companion star is insufficient to supply the accretion rate required for this luminosity. We therefore assume the BH accretes via Roche lobe overflow from a necessarily massive or giant companion. 

The IMBH interpretion depends on the X-ray emission from the disk in the immediate environment of the ULX, which is model dependent. Moreover, the physics at these high accretion rates near the event horizon is far from understood. We therefore seek an alternative channel to X-rays by which the nature of the ULX can be explored, using optical/infrared (optical/IR) properties. We argue that these properties will be strongly influenced by the proximity of such an intense X-ray radiation field, and this can be used as a diagnostic. If the ULX is in a binary system, the X-ray emission will modify the optical/IR characteristics of the companion star and accretion disk. In particular, this will induce intensity and colour shifts compared to normal stars, and these will vary at orbital periods. These both identify the true optical counterpart and also provide relatively direct indications of the masses of the components.

X-ray irradiation can drive evolution in XRB \citep{Podsi91,Ruder89}, and cause significant colour and magnitude changes of the optical counterpart. This has been observed in the sub-Eddington regime. One example is the Her X-1 system, an X-ray binary consisting of a neutron star accreting matter from a non-degenerate stellar companion. The X-ray luminosity is a third of the Eddington luminosity \citep{Howarth83}, and the binary period is $1.7$ days. The neutron star accretes matter via Roche lobe overflow through an accretion disk \citep{Vrtilek01}, and so is a good analogy to the ULX systems described in this paper. The star has been observed to change spectral type from A to B over the binary period. \citet{Bahcall72} observed a B magnitude amplitude of $1.5$ mag and interpreted this variation as a result of X-ray heating of a late A-type star. Other authors interpret the variation in terms of heating of the star and a tilted, precessing accretion disk \citep{Gerend76,Howarth83}.

We consider a binary model for ULX to investigate the radiation effects. We calculate the optical/IR observables from the companion star and accretion disk and discuss the implications of the results. We apply the results to the NGC 4559 candidates  to either eliminate them as possibilities or constrain the parameters of any binary system of which they may be part. We also predict the IR properties for future observations.

We describe our model in Section \ref{sec:model}, and examine the results of this model in Section \ref{sec:results}. In Section \ref{sec:4559} we compare the model to observations of ULX X-7 in NGC 4559.

\section{MODEL}
\label{sec:model}

The accretion rate required by the brightest ULXs exceeds that which could be supplied by a stellar wind. We therefore assume the matter is transferred onto the compact object through Roche lobe overflow. We will use this assumption to constrain the geometry of the system. We assume that the system is in a quasi-steady state, and the irradiated surfaces are in thermal, radiative, and hydrostatic equilibrium. This requires that the irradiated layers necessarily re-emit all of the radiation falling on them.

\subsection{The radiative transfer formulation}

We consider the effects of radiative transport and radiative equilibrium in the irradiated surface under X-ray illumination. We consider a plane-parallel model and adopt the radiative transport formulation of \citet{Milne26} and \citet{Wu01} to describe the heated stellar surface and accretion disk. We detail this formulation in Appendix \ref{sec:ppmodel}.

\subsubsection{The heated stellar surface}
\label{sec:starmodel}

The heating of any point on the stellar surface is described by the
equations in Appendix \ref{sec:ppmodel}. We calculated the total heating by
dividing the surface of the star into discrete cells and calculating the
magnitude of the effect for each cell. The Roche surface for a given mass ratio
was calculated over a grid of cells. Each cell has a flat surface, the size of which is dependent on the distance to neighbouring points. The angles $\alpha$ and $\theta$ (required components of the heating equations) can be calculated through appeal to the angle between the normal vector to the surface at the point in question, and the vector incident on the point originating at the BH ($\alpha$) or the observer ($\theta$). The irradiation temperature is calculated at this point, and taken to be the temperature over the entire surface of the
cell.

The shape of the Roche Lobe is determined solely by the mass ratio $q$, so the
angles and relative positions of the points are independent of the scale of the
system. The system can therefore be easily scaled appropriately to calculate the heating, according to the binary separation $a$. This in turn is directly inferred from the binary period when the mass ratio is known. If the system is semi-detached then the size of the star itself is also determined by the scale of the system.

We calculate the luminosity of the star by summing the emergent radiation in the direction of the observer for each cell. This includes a component as a result of irradiation, as well as the original stellar luminosity, which includes both limb darkening and gravity darkening \citep{vZeipel24} effects. 

\subsubsection{The heated accretion disk - Dubus et al.\  prescription}
\label{sec:diskmodel}
We also include the emission from an irradiated accretion disk as an additional optical source. 
We have considered two models. 
The first follows \citet{Dubus99} to describe this disk. 
There, the irradiation temperature $T_{irr}$ varies as 
\begin{equation}
T^4_{irr} = C {{\dot M c^2} \over {4 \pi \sigma R^2}}
\label{eqn:dubtemp}
\end{equation}
where $\dot M$ is the accretion rate and $R$ is the distance from the accreting source. 

For accreting black holes, 
\begin{equation}
L_x = \eta \dot M c^2,
\label{eqn:accrate}
\end{equation}
where the efficiency parameter $\eta \sim 0.1$.  
If we take an X-ray albedo of $0.9$ and assume that $\eta = 0.1$, 
  then the value of $C$ in Equation \ref{eqn:dubtemp} would be $\approx 2.57 \times 10^{-3}$, 
  with the geometry of a thin disk \citep{Dubus99,deJong96}. 
Here and hereafter we use Equation \ref{eqn:accrate} with $\eta = 0.1$ 
  to determine the value of $\dot M$ from $L_x$.

Using the fact that the radiation-transfer equations are linear, we can use the principle of superposition to calculate the disk temperature from the combination of the irradiative heating and the viscous heating in the disk in the absence of X-ray irradiation. We calculate the temperature as a function of disk radius, and then sum the flux from a series of blackbody annuli to describe the overall disk flux. We take the inner disk radius to be the last stable circular orbit around the BH we are describing. We take the outer disk radius to be the `tidal truncation radius', beyond which Keplerian orbits intersect. This is weakly dependent on the mass ratio \citep{Paczynski77} but is generally taken to be between $0.6$ and $0.7$ of the Roche lobe radius. We consider it sufficient in our model to fix the outer disk radius to be $0.6$ of the Roche lobe radius. 

\subsubsection{The heated accretion disk - radiative transfer formulation}
\label{sec:wudiskmodel}
The alternative description of the accretion disk directly  
  applies the radiative transfer formulation of \citet{Wu01} 
  that we have modified for for the star (see Appendix \ref{sec:ppmodel}).  
We consider a thin disk and determine a radial temperature profile in absence of irradiation 
  using the \citet{Shak73} prescription.  
We assume the local flare angle is given by $h(r) \propto r^{9/7}$ \citep{Dubus99}, 
  where $h$ is the disk scale height.

We calculate the luminosity of the disk in a manner identical to that of the star. We divide the disk surface into cells and calculate the heating effect on each as determined by the incident flux and the angle of incidence $\alpha$. We sum the emergent radiation from each cell in the direction of the observer in order to calculate the disk luminosity.

We will refer to both disk models throughout this paper. The prescription of \citet{Dubus99} will be referred to as the first model, and the radiative transfer formulation will be referred to as the second.

\section{MODEL RESULTS}
\label{sec:results}

\begin{table}
\centering
\begin{tabular}{@{}lcccccc}
\hline
Class & $log(M/$\Msun$)$ & $log(R/$\Rsun$)$ & $log(L/$\Lsun$)$\\
\hline
O5 V & 1.6 & 1.25 & 5.7\\
B0 V & 1.25 & 0.87 & 4.3\\
B5 V & 0.81 & 0.58 & 2.9\\
F0 I & 1.1 & 1.8 & 3.9\\
G0 I & 1.0 & 2.0 & 3.8\\
K0 I & 1.1 & 2.3 & 3.9\\
M0 I & 1.2 & 2.7 & 4.5\\
\hline
\end{tabular}
\caption{The set of parameters used in sections \ref{sec:results} and \ref{sec:4559} to describe stars of different spectral type and luminosity class. Values are taken from \protect \citet{Allen73}}
\label{tab:stars}
\end{table}

In this section we examine the dependence of our model on various parameters. 
Table \ref{tab:stars} contains masses, radii and luminosities 
for some early-type main sequence (MS) stars and some supergiants. 
We use these stellar parameters in our model. 
We keep the primary (BH) mass as an input parameter, and we constrain the scale of the Roche lobes 
by setting the volume radius of the secondary lobe equal to the radius of the star in Table \ref{tab:stars}, 
hence fulfilling the condition for Roche lobe overflow. 

We take the X-ray luminosity of the BH to be a constant $10^{40}$\ergss\ emitted isotropically. 
The soft X-rays are easily absorbed at the disk surface, 
whereas the hard X-rays are less easily absorbed but are scattered.  
They penetrate the star to greater optical depths 
  until the photons are down scattered to lower energies. 
In our model, 
  we parametrise the absorption coefficients for the hard and soft X-rays 
  by means of two parameters $k_s$ and $k_h$ 
  and denote the hard and soft components of the X-ray flux 
  as $S_h$ and $S_s$ respectively.   
We also define a hardness-ratio parameter $\xi=S_h/S_s$.    
We choose values of $2.5$ and $0.01$ for the two parameters $k_s$ and $k_h$ respectively throughout  
  and allow the band boundary of the hard and soft X-ray bands as parameters to be determined.  
For an input spectrum consisting of a blackbody and a power law component, 
  we find the boundary of the soft and hard band to be $1.5$keV.  
We set the gravity darkening parameter $\beta$ to be $0.25$, 
  representing a star with a purely radiative outer envelope.

\begin{figure*}
\centering
\begin{minipage}[c]{0.5\textwidth}
\includegraphics[width=1.0\textwidth]{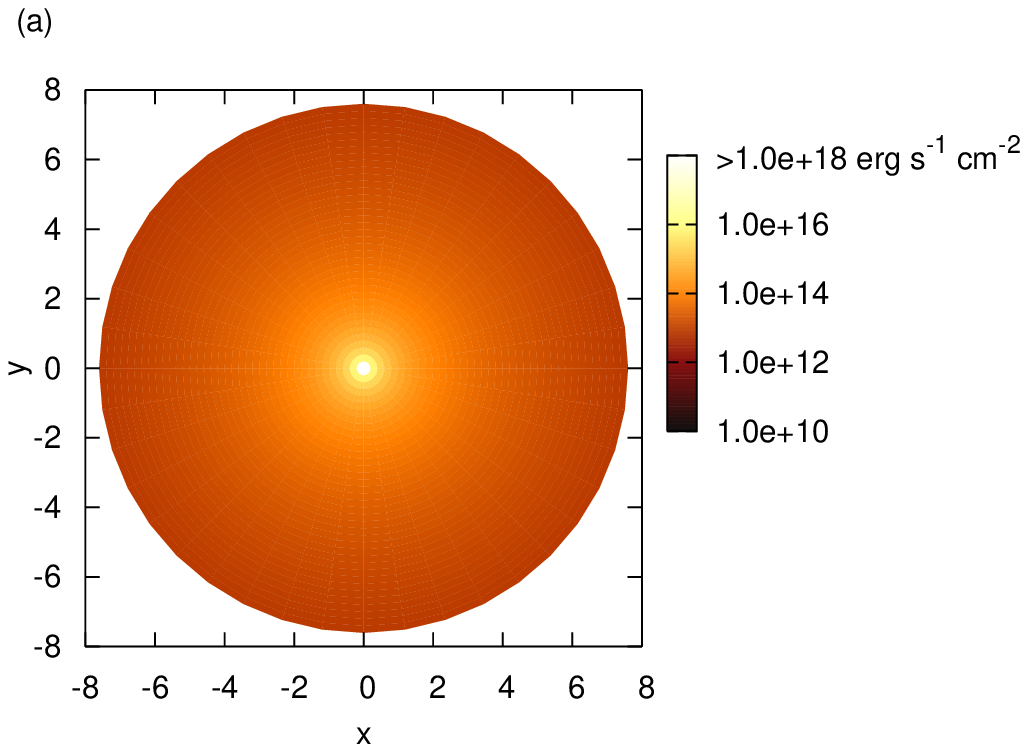}
\end{minipage}%
\begin{minipage}[c]{0.5\textwidth}
\hfill
\includegraphics[width=1.0\textwidth]{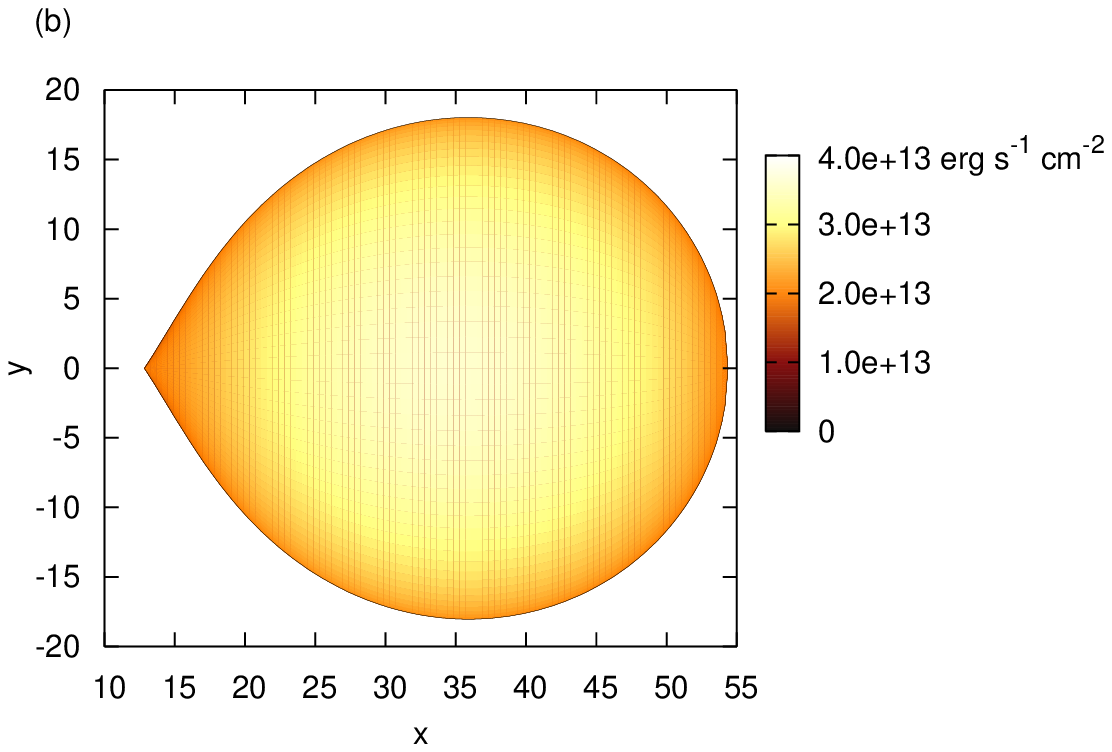}
\end{minipage}
\caption{The variation in intensity $B(\tau)$ with $\tau=2/3$ for (b) an irradiated O5V star and (a) a disk using the \protect \citet{Dubus99} prescription with $\xi=0.01$ and a BH mass of 10\Msun. We plot projections in the orbital plane with the labelled distances in units of \Rsun.}
\label{fig:intmap_ms}
\end{figure*}

\begin{figure*}
\centering
\begin{minipage}[c]{0.5\textwidth}
\includegraphics[width=1.0\textwidth]{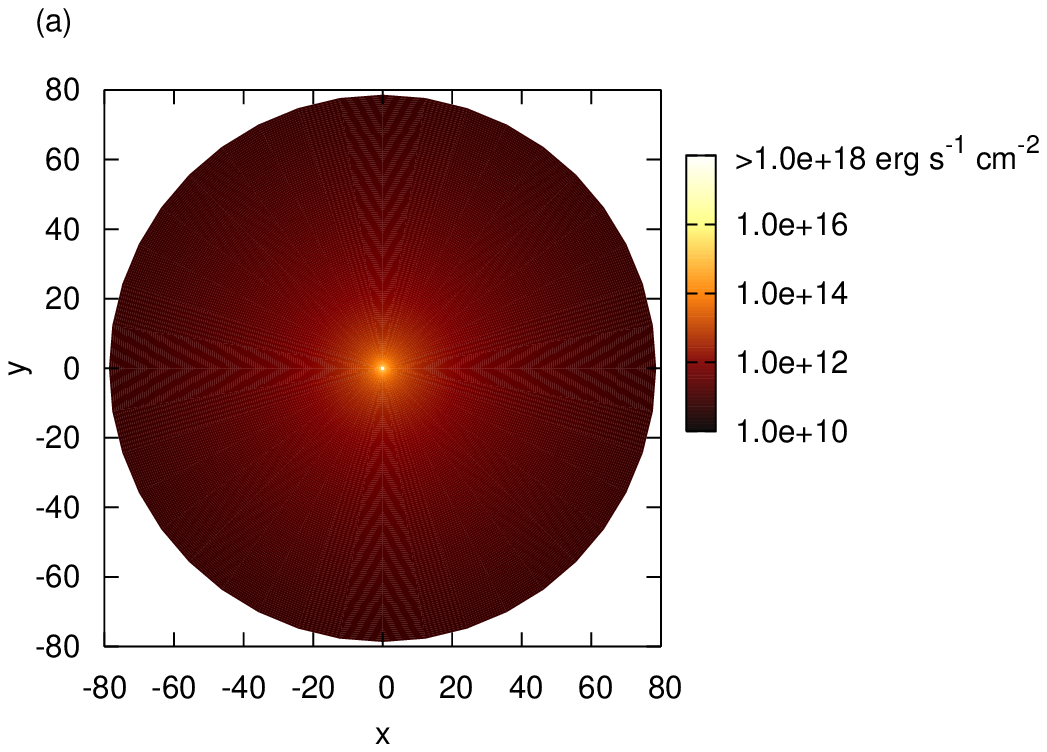}
\end{minipage}%
\begin{minipage}[c]{0.5\textwidth}
\hfill
\includegraphics[width=1.0\textwidth]{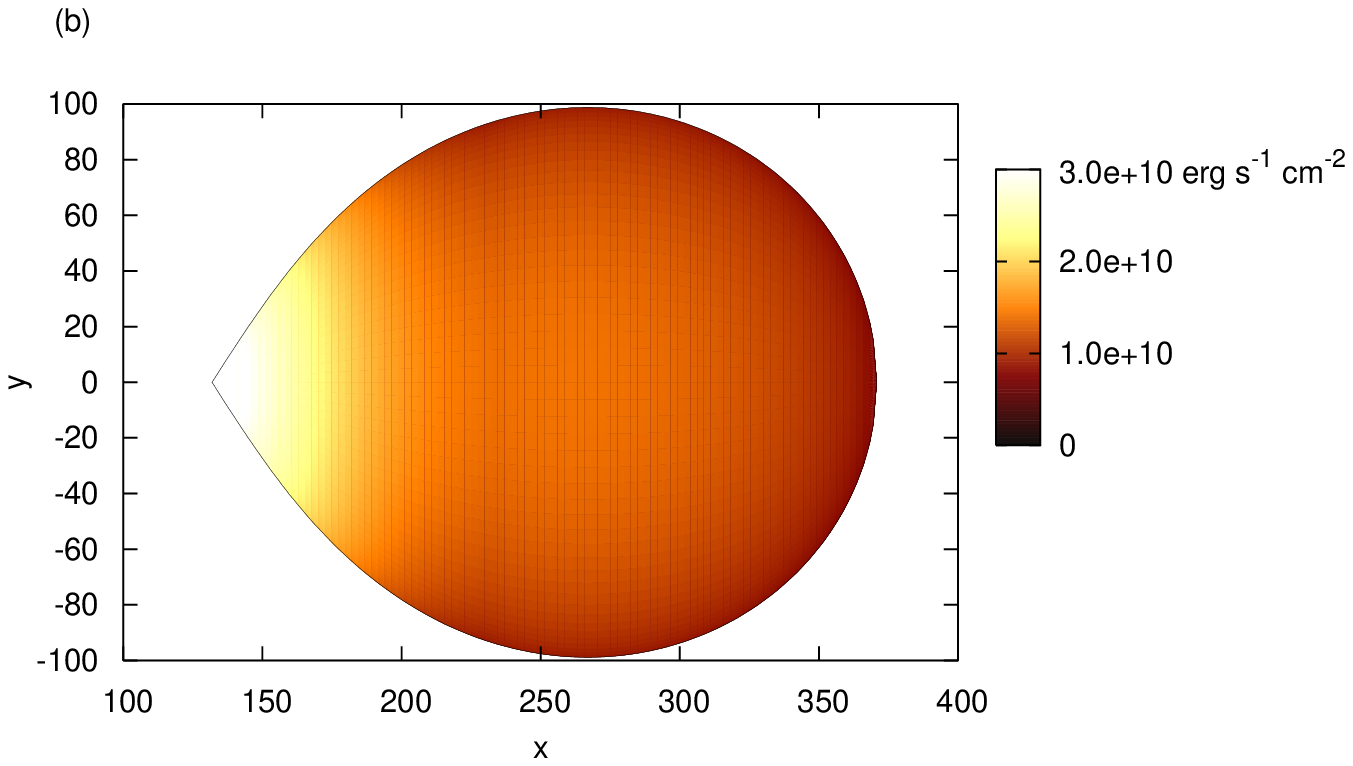}
\end{minipage}
\caption{The variation in intensity $B(\tau)$ with $\tau=2/3$ for (b) an irradiated G0I star and (a) a disk using the \protect \citet{Dubus99} prescription with $\xi=0.01$ and a BH mass of 10\Msun. We plot projections in the orbital plane with the labelled distances in units of \Rsun.}
\label{fig:intmap_sg}
\end{figure*}

In Figure \ref{fig:intmap_ms} we show the intensity variation over the surface of an O5V star and disk when we take the BH mass to be $10$\Msun. In Figure \ref{fig:intmap_sg} we use the same BH mass with a G0I star. We use the quantity $B(\tau)$ as a measure of intensity (equation \ref{eqn:btau}), setting $\tau$ to $2/3$, and we show projections of the star and disk in the orbital plane. The stellar maps show both the irradiative and darkening effects. We use the Dubus prescription of Section \ref{sec:diskmodel} to describe the disk. 

We see that the combined surface intensity is significantly higher than would be expected for an unirradiated star. There is however a noticeable difference between the two figures. The stellar intensity of the G0I star increases in the direction of the L1 point, reaching a peak there. On the other hand, in the O5V figure the darkening effects dominate at the L1 point, so that the intensity at that point is less than the surrounding surface. Here we are using a low hardness ratio of $\xi=0.01$, and so little flux penetrates to an optical depth of $\tau=2/3$. We find that if we increase the hardness ratio the intensity distribution becomes similar to that of the G0I star. As the BH mass increases, the separation increases, the irradiating flux decreases and the intensity distribution over the surface of both stars tends towards that shown in Figure \ref{fig:intmap_ms}(b). Note that we have not included here any shadowing of the accretion disk on the stellar surface, which should magnify any darkening at the L1 point.

\begin{figure*}
\begin{minipage}[c]{0.5\textwidth}
\includegraphics[width=1.0\textwidth]{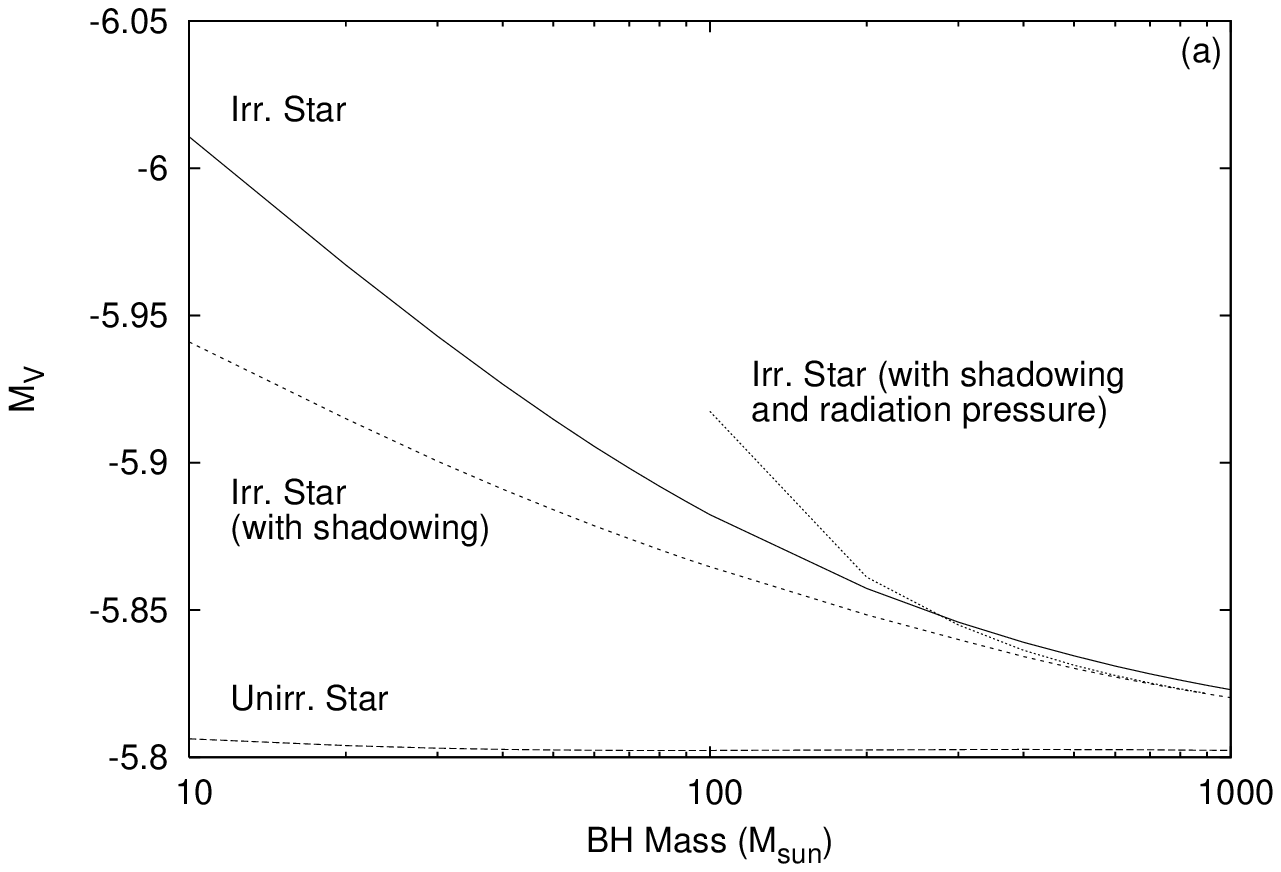}
\end{minipage}%
\hfill
\begin{minipage}[c]{0.5\textwidth}
\includegraphics[width=1.0\textwidth]{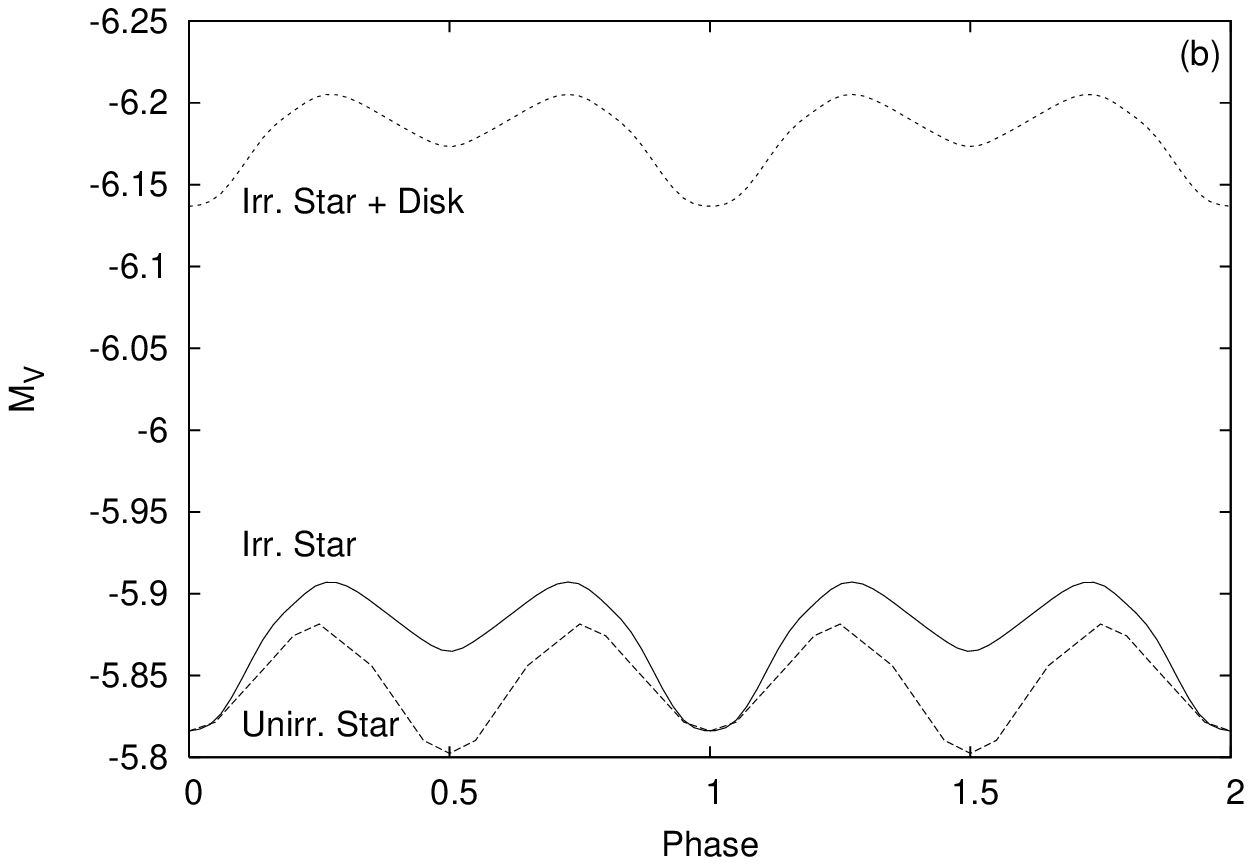}
\end{minipage}
\caption{(a) The effect of irradiation on an O5V star as a function of BH mass. We take $\cos i=0.5$, the X-ray luminosity to be $10^{40}$\ergss \ emitted isotropically and the star to be at superior conjunction. The hardness ratio $\xi$  is set to 0.01. We plot lines for an unirradiated star, an irradiated star using our model and an irradiated star when we consider the accretion disk to be shadowed on the stellar surface. We also plot a line for $100$\Msun \ $\leq$ BH mass $\leq$ $1000$\Msun, in which we include the effect of radiation pressure (Section \ref{sec:radp}). (b) The $V$ band absolute magnitude of the O5V star with binary orbital phase. We set the BH mass to $100$\Msun, $\xi$ to 0.01 and $\cos i=0.5$. We show the magnitude variation for both an unirradiated star and for a star irradiated by a source of $L_x=10^{40}$\ergss \ emitted isotropically. We also show the variation when an irradiated accretion disk is included.}
\label{fig:models}
\end{figure*}

In Figure \ref{fig:models}(a) we illustrate the change in effective luminosity of an O5V star. We show the $V$ band absolute magnitude against the BH mass for an unirradiated star and three different sets of irradiated star calculations. As well as our model described above, we shows the results from an extension to our model where we have included the effect of a disk which is completely opaque to the radiation incident on it and is hence shadowed on the stellar surface. We also show the effects of in addition taking into account the irradiation pressure on the star -- we will discuss this in section \ref{sec:radp}. The disk height at the outer disk radius $R_{out}$ is taken to be $0.2 R_{out}$ \citep{deJong96}. This shielding effect results in a reduced stellar magnitude. In this figure we have set the phase angle to be zero (so the star is in superior conjunction) and set the inclination of the system such that $\cos i = 0.5$. 

Figure \ref{fig:models}(a) shows that the heating effect on the star decreases with increasing BH mass, which may be counter-intuitive. This relationship is a consequence of constraining the volume radius of the secondary Roche lobe to the radius of the undistorted star. As the mass ratio decreases, the Roche lobe geometry requires the binary separation $a$ to increase. The result is a decrease in the amount of flux incident on the stellar surface.

Figure \ref{fig:models}(b) is a sample lightcurve for the O5V star. Here we use a BH mass of $100$\Msun \ and an inclination such that $\cos i = 0.5$. This figure shows both the ellipsoidal variation of an unirradiated star, as well as the combination of both ellipsoidal and irradiative effects. We include a third line showing the magnitude when the irradiated accretion disk is included. We will discuss the effect of the disk on the lightcurve in Section \ref{sec:diskresults}. 

\subsection{Irradiation pressure effects}
\label{sec:radp}

We now consider the effect of the X-ray irradiation on the geometrical shape of the secondary. We apply the prescription of \citet{Phillips02}, which involves a modification of the Roche potential in which the radiation pressure force is parameterised using the ratio of the radiation to the gravitational force. Equation 6 of \citet{Phillips02} shows how we can combine the gravitational and radiation pressure forces from the BH as a 'reduced' gravitational force equal to $(1-\delta)F_{grav}$, where $F_{grav}$ is the gravitational force. $\delta$ is the product of a parameter dependent on the X-ray luminosity and binary mass ratio (calculated using equation 25 of \citealt{Phillips02}) and the cosine of the angle between the surface normal and the direction of the flux vector.

After calculating the position of each cell on the Roche surface in our usual way, we calculate a $\delta$ value for each cell, replace the gravitational force in the Roche potential with the reduced gravitational force and recalculate the position of each cell. This needs to be repeated for a number of iterations in order to find the solution since the surface normal for each cell changes for each calculation. When the surface has been determined we calculate the heating at each point in our normal way. We take the accretion disk to be opaque to the radiation, so the inner Lagrangian point is shadowed and therefore the radiation pressure will not cause the star to become detached from this point. 

\begin{figure}
\begin{minipage}[c]{0.45\textwidth}
\includegraphics[width=1.0\textwidth]{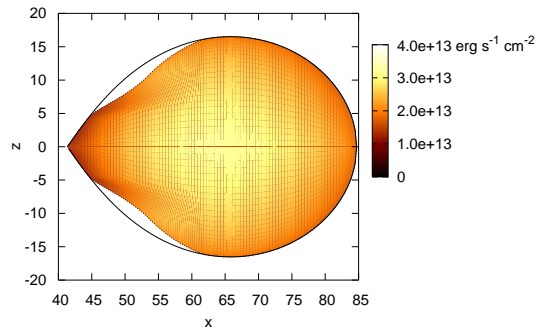}
\end{minipage}
\caption{The variation in intensity $B(\tau)$ with $\tau=2/3$ for an irradiated O5V star with a BH mass of 150\Msun \ and with disk shadowing. We include the formulation of \protect \citet{Phillips02} in order to show the effect of irradiation pressure on the star. The solid outline shows the stellar shape when we neglect this effect. We plot a projection perpendicular to the orbital plane with the labelled distances in units of \Rsun.}
\label{fig:intmap_rp}
\end{figure}

The greater the flux incident on the stellar surface, the greater the pressure effect. We therefore see the most significant distortion when the binary separation is at its lowest, which occurs when we use a MS star and a low BH mass. We illustrate this effect in Figure \ref{fig:intmap_rp}, in which we use a $150$\Msun \ BH and an O5V star. We show the $V$ magnitude dependence on increasing BH mass in Figure \ref{fig:models}(a). 

Note that we plot values only for BH masses of $100$ -- $1000$\Msun. This is because we observe that for a BH mass of less than $100$\Msun \ the flux incident on the surface is extremely high and the \citet{Phillips02} formulation becomes inappropriate to describe the stellar shape, since the formulation does not allow for any surface motion. In reality the external irradiation will drive circulatory currents in the stellar surface. A full treatment will require hydrodynamical motions to be considered.

\subsection{Inclusion of the accretion disk}
\label{sec:diskresults}

We now investigate further the additional flux from the accretion disk. An increased BH mass leads to a larger binary separation and thus to a corresponding increase in the size of the accretion disk, since the outer disk radius is constrained by the Roche lobe size through tidal effects. The net result is that the disk total luminosity increases with BH mass, and hence compensates for the decreasing stellar total luminosity. 

\begin{figure*}
\centering

\begin{minipage}[c]{0.5\textwidth}
\includegraphics[width=1.0\textwidth]{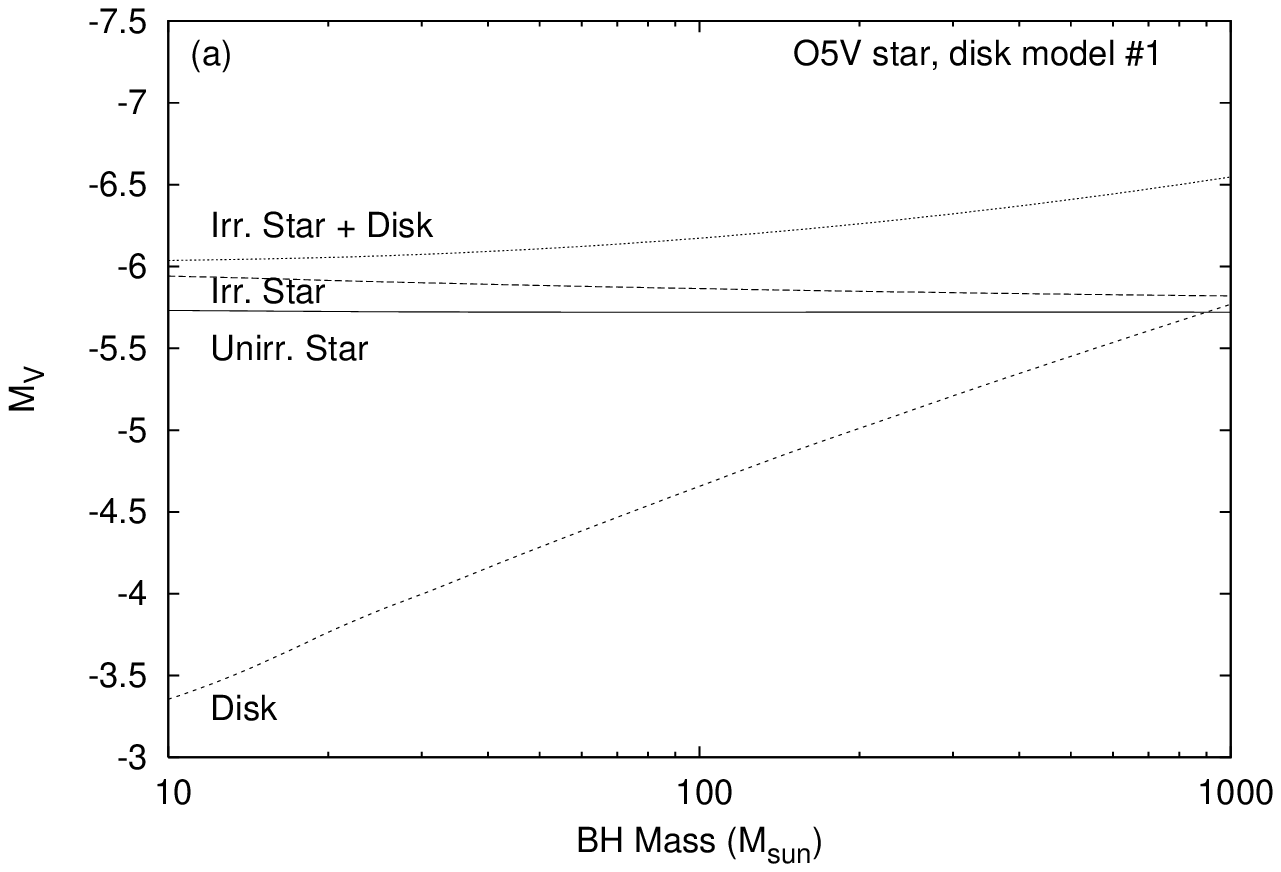}
\end{minipage}%
\begin{minipage}[c]{0.5\textwidth}
\hfill \includegraphics[width=1.0\textwidth]{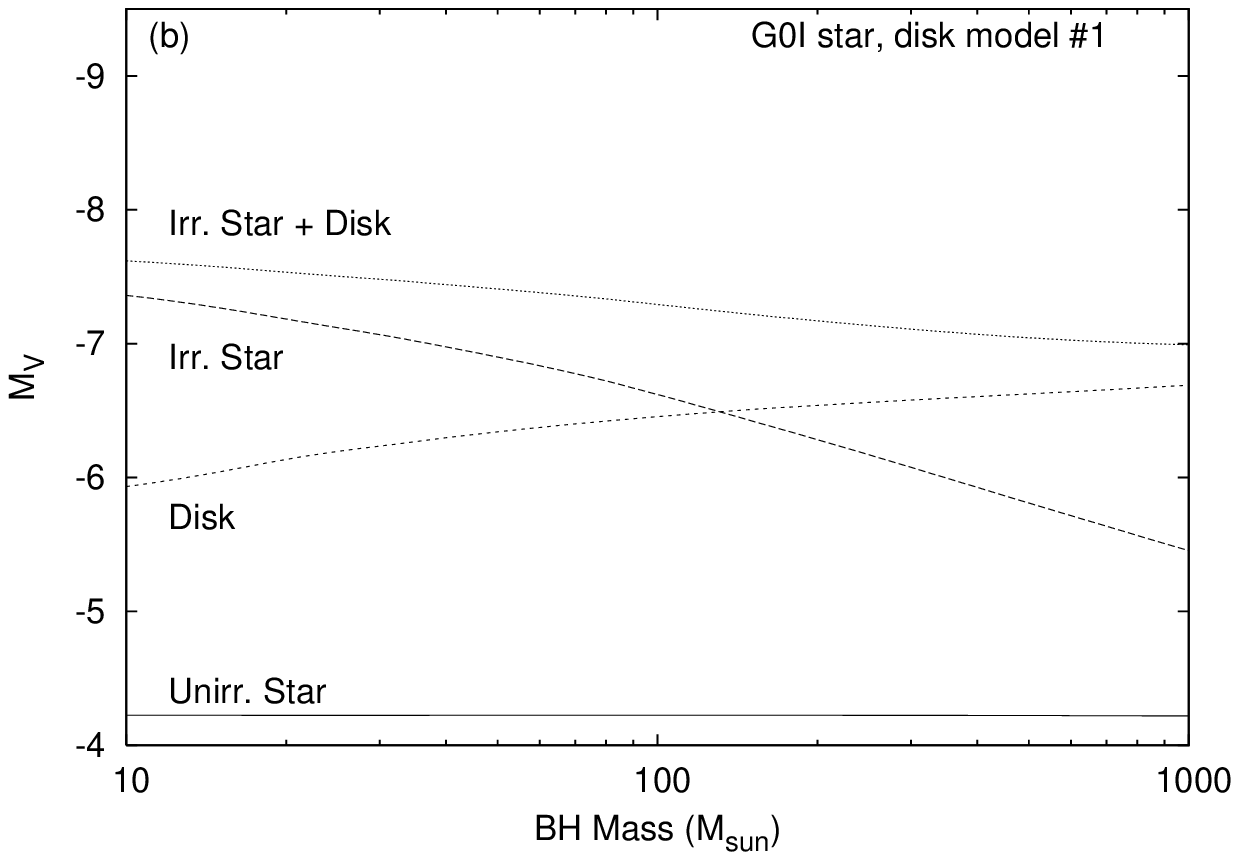}
\end{minipage}

\begin{minipage}[c]{0.5\textwidth}
\includegraphics[width=1.0\textwidth]{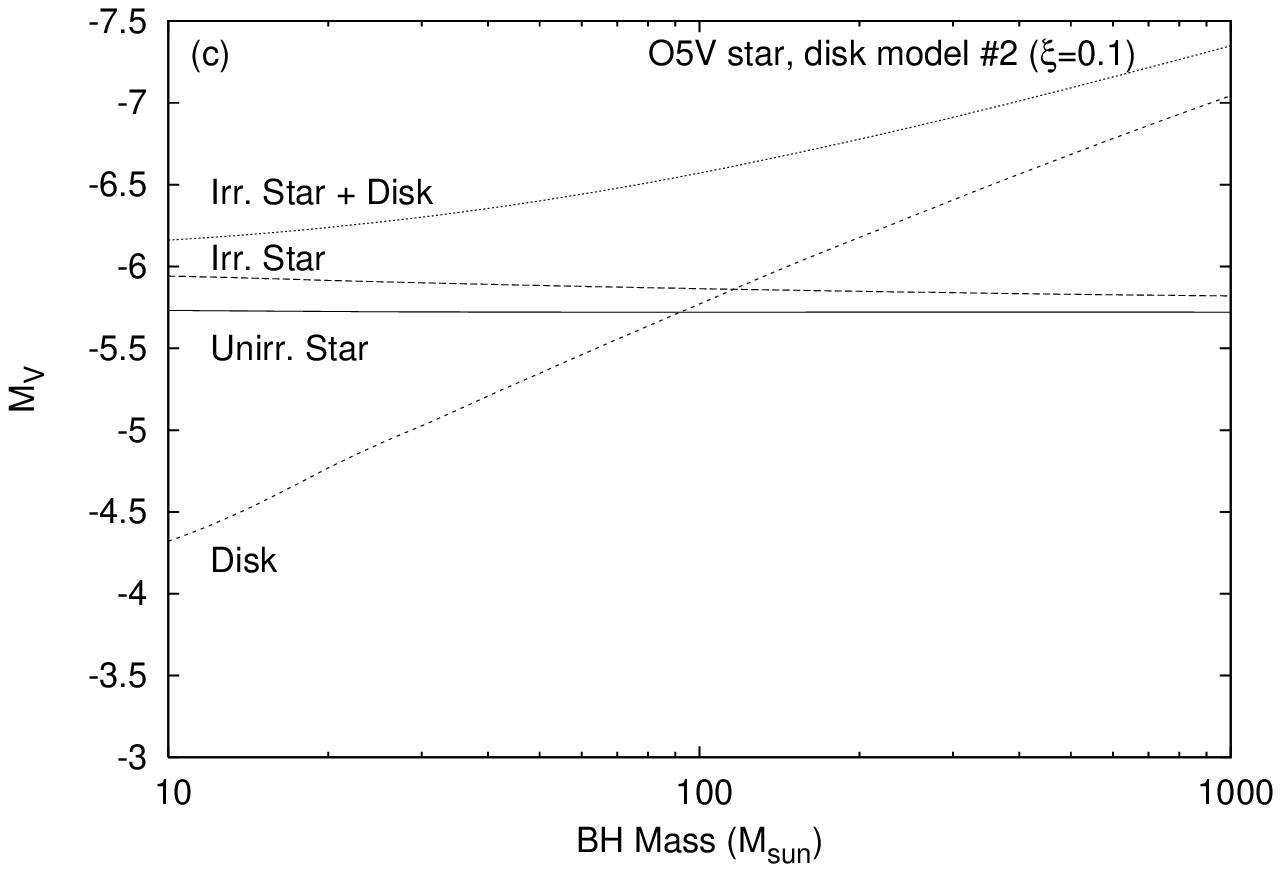}
\end{minipage}%
\begin{minipage}[c]{0.5\textwidth}
\hfill \includegraphics[width=1.0\textwidth]{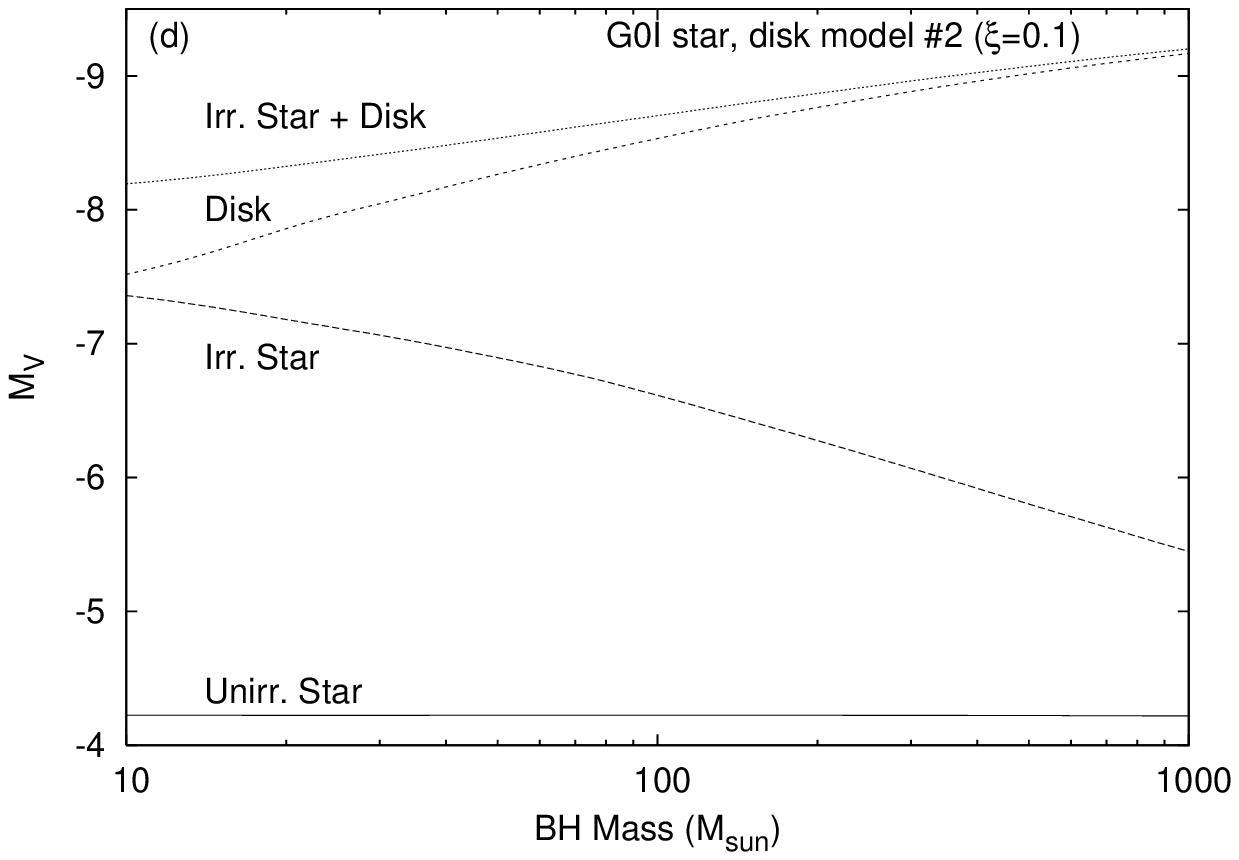}
\end{minipage}

\caption{$V$ band absolute magnitudes for irradiated stars and accretion disks. We set $L_x=10^{40}$\ergss, $\cos(i)=0.5$ and take the star to be at superior conjunction. We use two different stars in our model -- an O5 type MS star (a,c) and a G0 type supergiant (b,d). We plot the visual magnitude of the star and disk against BH mass using the \protect \citet{Dubus99} disk formulation of Section \ref{sec:diskmodel} (a,b) and the \protect \citet{Wu01} disk formulation of Section \ref{sec:wudiskmodel} (c,d). The disk magnitude in this second formulation is very dependent on the hardness ratio $\xi$ of the X-ray illuminating flux (see Appendix \ref{sec:ppmodel}). We use $\xi=0.1$, at which the disk magnitude is greater than for the first prescription.}
\label{fig:massmags}
\end{figure*}

\begin{figure*}
\centering
\begin{minipage}[c]{0.5\textwidth}
\includegraphics[width=1.0\textwidth]{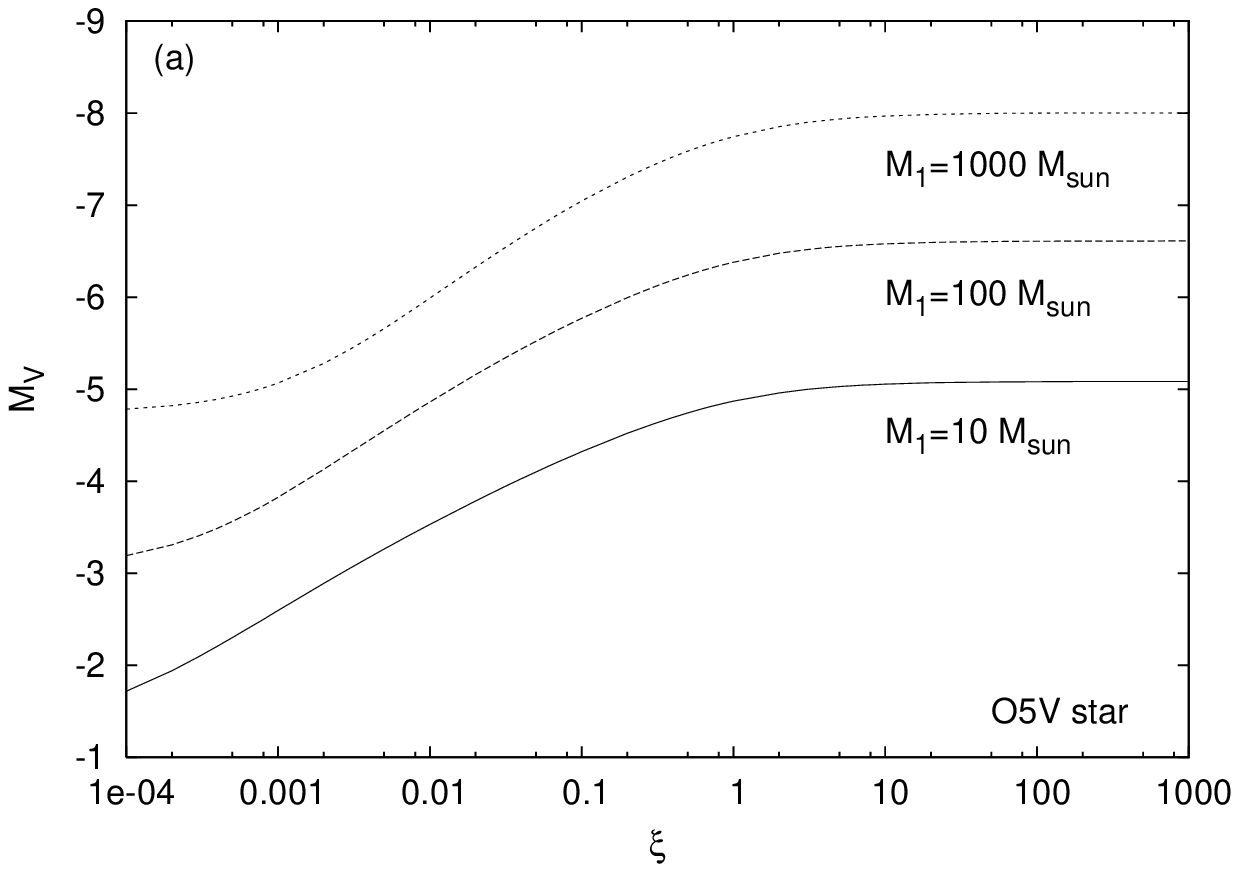}
\end{minipage}%
\begin{minipage}[c]{0.5\textwidth}
\hfill \includegraphics[width=1.0\textwidth]{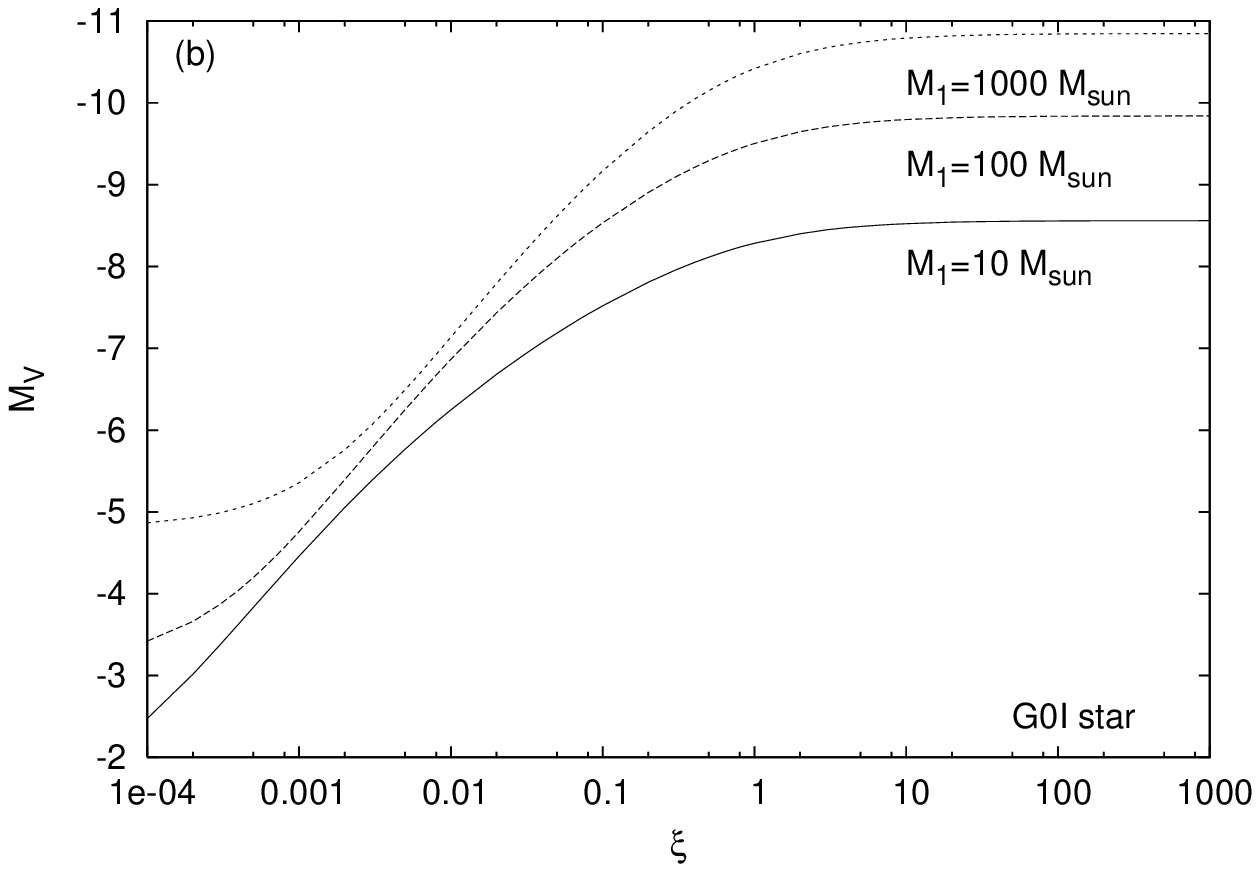}
\end{minipage}

\caption{The $V$ band absolute magnitude of an irradiated accretion disk using the \protect \citet{Wu01} disk formulation for different values of the hardness ratio $\xi$. We plot values for BH masses of $10$, $100$ and $1000$\Msun \ with an O5 type MS star (a) and a G0 type supergiant (b). We set $L_x=10^{40}$\ergss and $\cos(i)=0.5$.}
\label{fig:chiplots}
\end{figure*}

We find that the luminosity class of the irradiated star is the most important factor in determining which component dominates. To illustrate this we show in Figure \ref{fig:massmags} the absolute magnitude dependence on BH mass for a O5V and a G0I star, along with the corresponding disk magnitudes. We use the \citet{Dubus99} disk prescription for Figure \ref{fig:massmags}(a,b), and the \citet{Wu01} disk model for Figure \ref{fig:massmags}(c,d). We find the magnitude of this second disk model is strongly dependent on the hardness parameter $\xi$. A low value of $\xi=0.01$ produces a disk almost identical to the Dubus disk in terms of magnitude and colour. We use a value of $\xi=0.1$ in (c) and (d) to illustrate the effect of a harder X-ray spectrum. In Figure \ref{fig:chiplots} we further illustrate the effect of varying disk hardness on the disk magnitude for different combinations of star and BH. 

If we examine the stellar luminosity change as a function of BH mass in Figure \ref{fig:massmags} first, we note that while the BH + MS star changes by a few tenths of a magnitude over the BH mass range, the BH + supergiant decreases by two magnitudes over that same range. The supergiant has a much larger radius, and so for a low binary separation the flux incident on the stellar surface will be high. However, when the mass ratio is decreased, this larger radius leads to a correspondingly larger binary separation than we see in the MS systems.

If we now examine the disk intensity dependence on BH mass, we find the reverse is true. If the donor is a MS star, the Dubus disk (a) increases in $V$ magnitude by more than 2.5 magnitudes over the mass range. When we use harder X-ray radiation (c) the result is a more luminous disk, with approximately the same increase in magnitude over the mass range. In contrast, the Dubus disk accompanying the supergiant (b) increases in $V$ magnitude by less than a magnitude over the mass range (the exact increase depends on the method we use to determine $R_{out}$). The disk irradiated by the harder X-rays  (d) increases by about 1.5 magnitudes. These can be explained by the fact that the  large supergiant leads to a large Roche lobe for all BH masses. Hence even a low BH mass results in a very bright disk, and since the temperature of the disk decreases with increasing disk radius, the effect of making a large disk larger still has a smaller effect in terms of total disk luminosity. In constrast, when the companion star is on the main sequence, the smaller size of the system at low BH masses results in a small and faint disk. When the BH mass is increased and the disk grows, the effect on its magnitude is much more significant.

We also show on Figure \ref{fig:massmags} the $V$ magnitude dependence on BH mass of the disk and star combined. It is interesting to note that were we actually observing an O5V system, it would be much easier to constrain the BH mass with the disk component included. The same cannot be said for the system with the G0 supergiant. The gradient of the luminosity change with increasing BH mass is still dictated by the decreasing stellar luminosity, but the curve is rendered shallower by the disk component.

We now investigate the effect of the accretion disk on the amplitude of the lightcurve. We include a line in Figure \ref{fig:models}(b) showing the magnitude when the irradiated accretion disk is included. In our thin disk approximation the contribution of the disk will be constant for any phase, so the shape of the lightcurve will not be affected. The exception to this will be when the inclination is such that the disk is partially or fully eclipsed by the star. The relative amplitude of the lightcurve will be affected, depending on the luminosity of the disk. In Figure \ref{fig:models}(b) for example, the amplitude decreases from $\sim 0.1$ V magnitudes to $\sim 0.07$. For any given star the magnitude of the disk will increase with BH mass, and so the lightcurve relative amplitude will decrease. The lightcurve of the star alone will decrease in amplitude with increasing BH mass due to the decrease in X-ray flux incident on the star -- the addition of the disk will reinforce this. We found that the amplitude of the lightcurve for a G0I star with $L_x=10^{40}$\ergss, $\cos(i)=0.5$ and $\xi=0.01$ drops from $\simeq 1.5$ Mag to $\simeq 0.2$ Mag as we increase the BH mass from $10$ to $1000$\Msun. An O5V star under the same set of conditions produces a lightcurve with an amplitude of $\simeq 0.17$ Mag for a BH mass of $10$\Msun. This drops below $0.1$ Mag as the mass is increased to $100$\Msun \ as shown in Figure \ref{fig:models}(b), and at a mass of $1000$\Msun \ the lightcurve is dominated by the ellipsoidal variation and has an amplitude of $\simeq 0.05$ Mag. 

Figure \ref{fig:chiplots} shows the disk magnitude for a hardness ratio over the range of $\xi=10^{-4}$ -- $10^4$ using the \citet{Wu01} formulation. We show the magnitude for a combination of a $10$, $100$ and $1000$\Msun \ BH with a O5V (a) and a G0I (b) star. The change in disk luminosity over this hardness range is large, demonstrating the importance of this factor.

To summarise this section, we find that the stellar luminosity component is at its greatest for low BH masses and the disk component is at its greatest for high BH masses. If we consider separately a MS star, a supergiant star, a disk in a BH/MS system and a disk in a BH/supergiant system, we find the biggest changes in magnitude over the BH mass range occur for a supergiant star or a BH/MS disk. In general then, while the emission will always consist of a disk and a star component, the stellar component will dominate for a MS star / low mass BH combination, and the disk component will dominate in the case of a supergiant / high mass BH. This assumes the X-ray radiation is soft -- when the hardness of the X-rays is increased, the contribution of the disk component will increase for all BH masses, and in the supergiant systems in particular we begin to see domination by the disk component over the entire mass range.

\subsection{Irradiation effects at infrared wavelengths}
\label{sec:IR}

We now use our model to examine the change in the parameters of a ULX system by making predictions at other wavelengths, extending into the infrared. 

\begin{figure}
\centering

\begin{minipage}[c]{0.5\textwidth}
\includegraphics[width=1.0\textwidth]{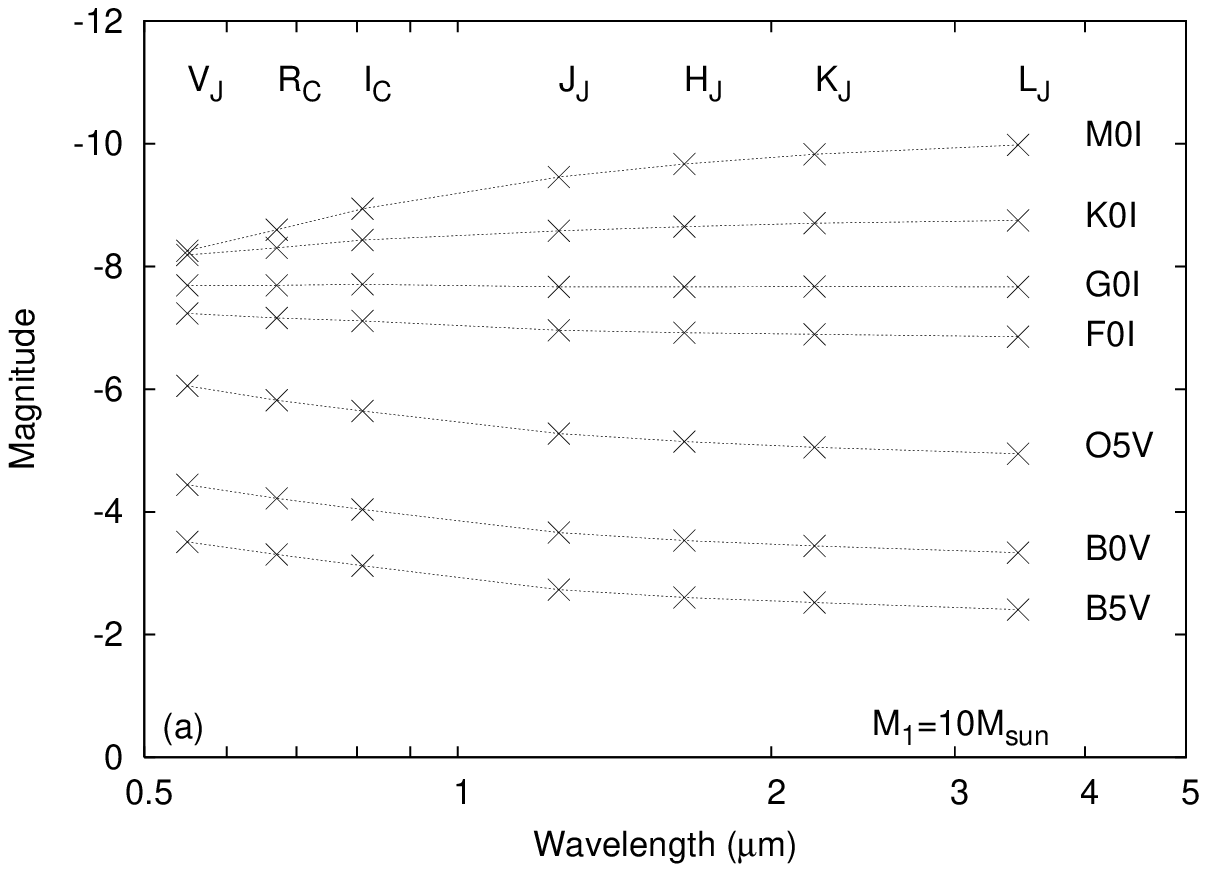}
\end{minipage}%

\begin{minipage}[c]{0.5\textwidth}
\includegraphics[width=1.0\textwidth]{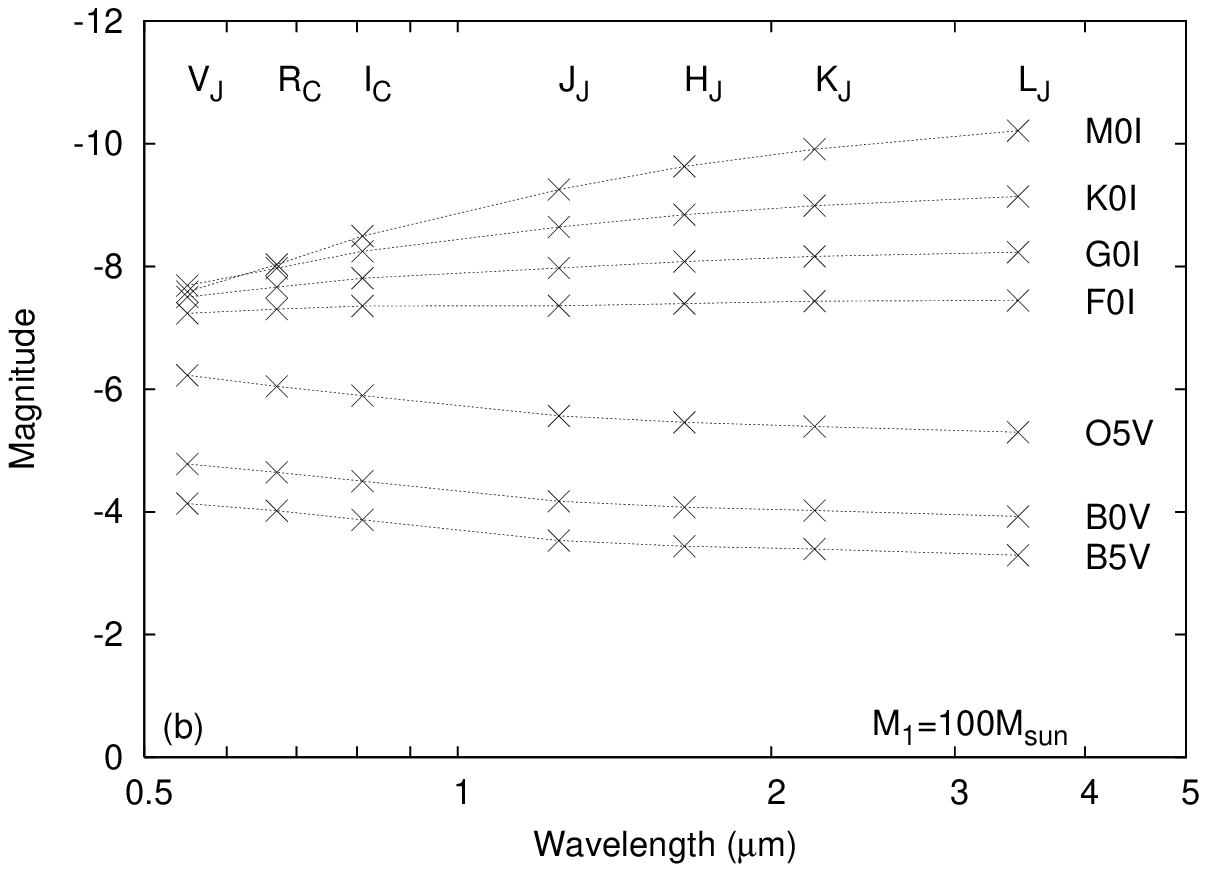}
\end{minipage}%

\begin{minipage}[c]{0.5\textwidth}
\includegraphics[width=1.0\textwidth]{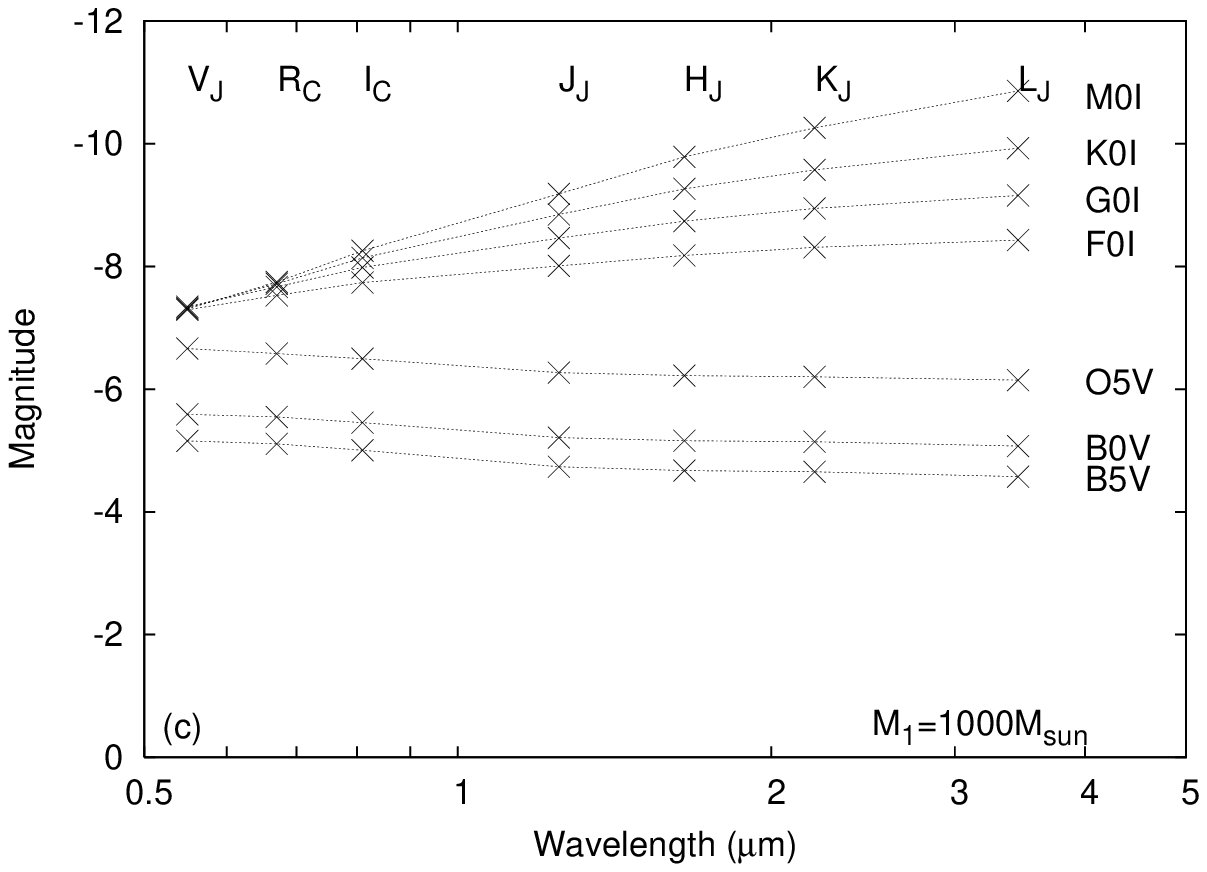}
\end{minipage}

\caption{The absolute magnitude of various irradiated stars and disks at wavelengths of $0.5 - 4.0 \mu$m. We take $\cos i=0.5$, the X-ray luminosity to be $10^{40}$\ergss \ and the star to be at superior conjunction. We show the magnitudes (from top to bottom) for BH masses of $10$, $100$ and $1000$\Msun. We use the \protect \citet{Wu01} disk prescription with $\xi=0.01$.}
\label{fig:freqs}
\end{figure}

We have examined the magnitude change for a star and disk for a wavelength of $0.5 - 4.0 \mu$m, encompassing the $V$, $R$, $I$, $J$, $H$, $K$ and $L$ wavebands. We use the Johnson filter convention, with the Kron/Cousins convention for the $R$ and $I$ bands. We show plots in Figure \ref{fig:freqs} for the stars in Table \ref{tab:stars}, using three different BH masses and the \citet{Wu01} disk model with $\xi=0.01$. As noted before, this is essentially interchangeable with the Dubus disk. We use a $10$\Msun \ BH in Figure \ref{fig:freqs}(a), a $100$\Msun \ BH in Figure \ref{fig:freqs}(b) and a $1000$\Msun \ BH in Figure \ref{fig:freqs}(c). We incorporate shadowing of the star by the disk into our stellar irradiation model.

Firstly, we see that there is a very large range in magnitude between these different systems. Secondly, we notice that as the mass of the BH increases, it becomes progressively harder to distinguish between different star/disk combinations with a $V$ band observation alone. Thirdly, we see that there is a much more clear distinction when we extend observations to longer wavelengths. Note that there is a clear separation between the MS stars and the supergiants which becomes more apparent as BH mass is increased. This suggests that infrared observations will have more diagnostic power in determining the characteristics of the ULX than observations at optical wavelengths. 

\section{APPLICATION TO ULX X-7 IN NGC 4559}
\label{sec:4559}

\citet{Soria05} used \hst \ data to study the optical environment of ULX X-7 in NGC 4559. They found eight possible candidates for the ULX optical counterpart, listing the $B$, $V$ and $I_{C}$ standard magnitudes for each in table 2 of that paper. In this section we apply our model to this system with the aim of further constraining the candidate population.

We use colour-magnitude diagrams to compare our model predictions with the observations of \citet{Soria05}. We again use the parameters of Table \ref{tab:stars}. This set is sufficient for us to make generalisations about the spectral type and luminosity class of the donor star. We also add to this set the parameters inferred in \citet{Soria05} for `Star No.1'; the most likely candidate for the optical counterpart. These are determined from the evolutionary tracks of \citet{Lejeune01} for a non-irradiated, isolated star with the observed $B$, $V$ and $I$ colours. The mass is found to be $15-30$\Msun (we take the mass to be in the middle of this range), the bolometric luminosity is $\approx 1.4 \pm 0.2 \times 10^5$\Lsun \ and the effective temperature is $T_{eff} = 16000 \pm 5000 K$. By additionally using these parameters we can make predictions for a slightly evolved star, based on a current evolutionary model.

In each colour-magnitude diagram we plot the colours and magnitudes of seven of the eight candidate stars. No data are available for star no.7, which fell on a hot \hst/WFPC2 pixel. The error on the colour measurement is taken from the errors on the individual magnitude measurements, with the assumption that these errors are not correlated between bands. We then plot a line for each set of stellar parameters which we used in our model. The line shows the effect of varying the BH mass. The mass is varied from $10-1000$\Msun, and we indicate on each line where the mass is equal to $10$, $100$ and $1000$\Msun.

\begin{figure}
\centering

\begin{minipage}[c]{0.5\textwidth}
\includegraphics[width=1.0\textwidth]{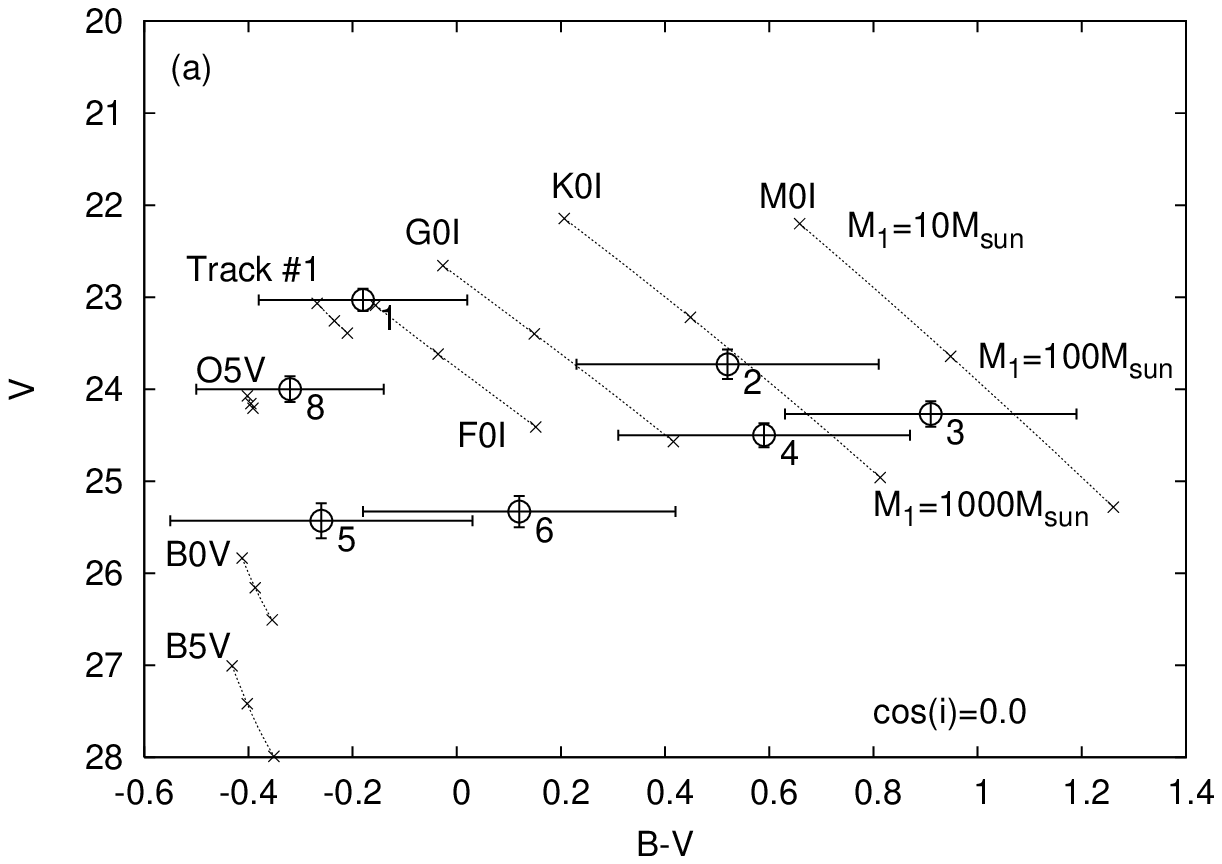}
\end{minipage}%

\begin{minipage}[c]{0.5\textwidth}
\includegraphics[width=1.0\textwidth]{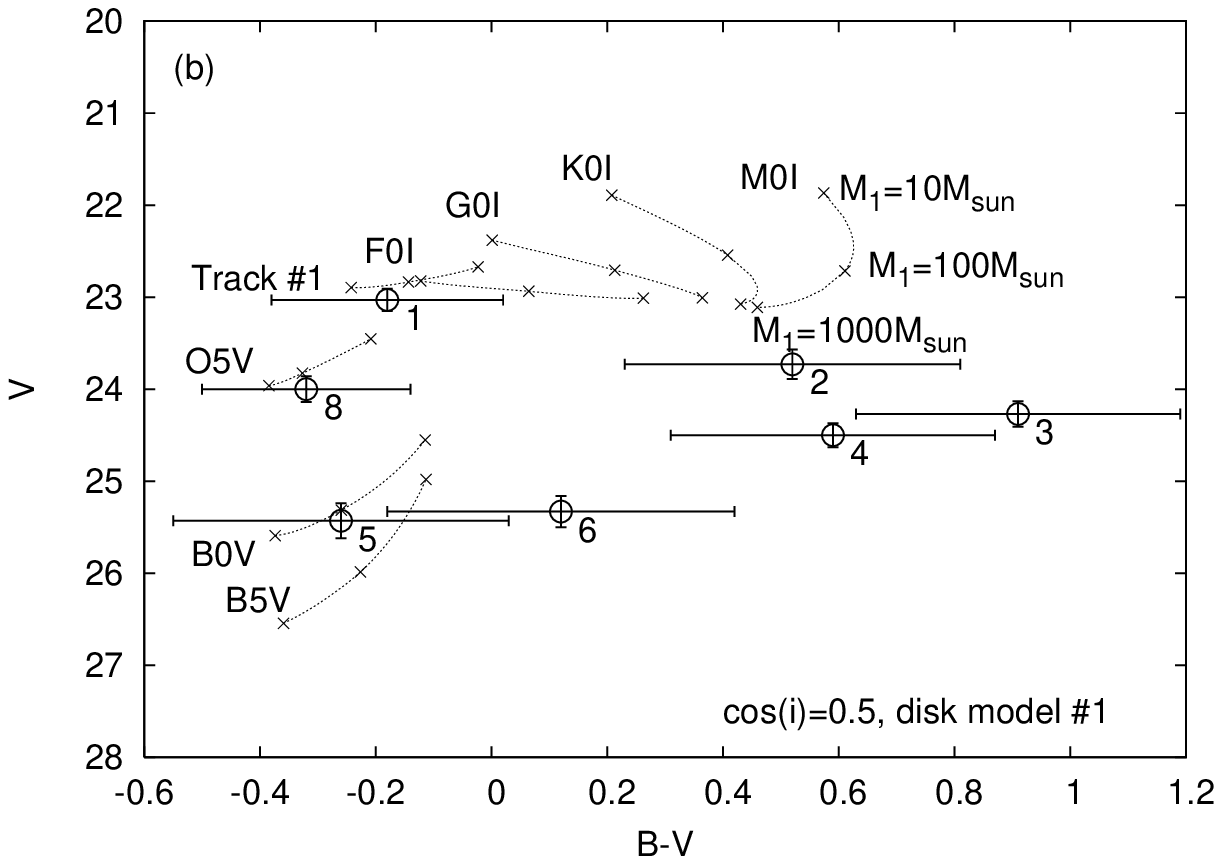}
\end{minipage}%

\begin{minipage}[c]{0.5\textwidth}
\includegraphics[width=1.0\textwidth]{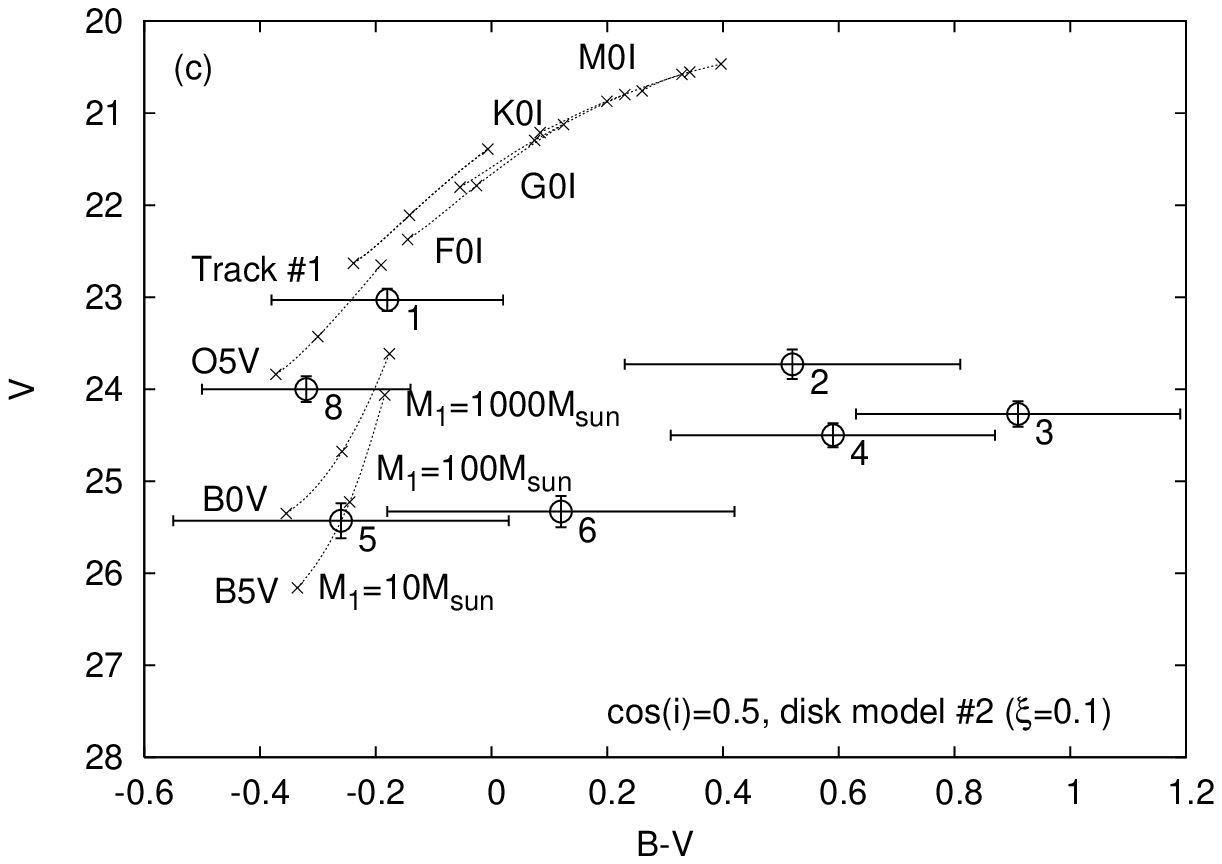}
\end{minipage}

\caption{Colour-magnitude diagram comparing observations of candidates for the counterpart of X-7 in NGC 4559, with model predictions for a range of different stars. We use the stars of Table \ref{tab:stars}, along with a star with parameters derived by applying the evolutionary model of \protect \citet{Lejeune01} to the observation of star no.1 \citep{Soria05}. This has been labelled `Track \#1'. For each star we plot a line showing the model result for a BH mass of $10-1000$\Msun. We set the X-ray luminosity to be $10^{40}$\ergss \ and take the star to be at superior conjunction. Figure (a) corresponds to an inclination of $\cos(i)=0.0$, so there is no disk component present in the emission. In (b) and (c) we set $\cos(i)=0.5$, using the \protect \citet{Dubus99} disc formulation in the former and the \protect \citet{Wu01} formulation with $\xi=0.1$ in the latter.}
\label{fig:4559}
\end{figure}

Figure \ref{fig:4559} compares the stars of Table \ref{tab:stars} with the observations for systems for different inclinations, since we have no knowledge of the orientation of the system. We take the inclination to be $\cos(i)=0.0$ in Figure \ref{fig:4559}(a) and $\cos(i)=0.5$ in Figure \ref{fig:4559}(b,c). As before we use two disk models -- the first of which describes both the prescription of Section \ref{sec:diskmodel} and the model of Section \ref{sec:wudiskmodel} when $\xi=0.01$. The second disk model uses the formulation of Section \ref{sec:wudiskmodel} but with $\xi=0.1$, giving a brighter disk. We have again taken the phase angle of the binary to be such that the star is at superior conjunction. We have examined other phase angles but we find that this does not affect the general results described in this section. We also find that the effect of including radiation pressure for BH masses of $100$ -- $1000$\Msun \ is small. We incorporate shadowing of the star by the disk into our stellar irradiation model.

We first examine Figure \ref{fig:4559}(a). This describes the case where $\cos(i)=0.0$, so we see no accretion disk component. Examining the lines for the Table \ref{tab:stars} stars, we see a clear distinction between the supergiants and the MS stars. The MS stars occupy the left hand side of the plot only, and show much less change in colour and magnitude over the BH mass range. If we now look at the observed stars, we can see most of the candidate stars seem to fit into one of the two regimes. It is possible that either star 5 or 8 could be an irradiated MS star. Stars 2, 3 and 4 could all be irradiated supergiants (albeit for a very large BH mass). Star 1 could be an irradiated supergiant, or an irradiate, evolved MS star (track 1).

The picture changes significantly when we consider the case where $\cos(i)=0.5$. As well as the irradiated star, there is a disk component present. We look first at the fainter disk as described in Figure \ref{fig:4559}(b). As in Figure \ref{fig:massmags}, we see that the inclusion of the disk results in an increase in $V$ magnitude with increasing BH mass. Stars 1, 5 and 8 can still be described by our models.. However, the addition of the disk flux has a large effect on the magnitude of the supergiant and high-mass BH systems. The large luminosity decrease with increasing BH mass is curtailed, and the observations of stars 2, 3 and 4 are too faint for supergiants to be candidates. We additionally examined the case where $\cos(i)=1.0$, and found the difference between this and the $\cos(i)=0.5$ case to be minor. Again, stars 1, 5 and 8 fit with predictions of a star and disk. Stars 2, 3 and 4 are too faint to fit with the predictions of the model.

We now consider the brighter disk caused by increasing $\xi$ by an order of magnitude, as described in Figure \ref{fig:4559}(c). This disk completely dominates the optical emission -- the only role the star plays is to constrain the disk brightness by determining the Roche lobe geometry. We draw the same conclusions from these plots as we did from Figure \ref{fig:4559}(b) -- Stars 1, 5 and 8 fit within the predictions of these star and disk combinations. Stars 2, 3 and 4 do not.

We find therefore that the observed colours of stars 2, 3 and 4 are marginally consistent with irradiated red supergiants when we assume a large BH mass and set $\cos(i)=0.0$. In this case there will be temporal variations which will act as the candidate signature. When we move to lower inclinations the increased flux from the accretion disk in our models indicates that the optical counterpart cannot be a red star (redder than $B-V \sim 0.6$), and so we suggest that star 1, 5 or 8 is more likely to be the counterpart. In this case, the optical observations are unable to discriminate between different BH masses. We note that \citet{Soria05} suggested without taking the effects of a disk or irradiated companion into account that star 1 is the most likely candidate for the optical counterpart of X-7 in NGC 4559. In addition, taking the results of Section \ref{sec:IR} into account, we note that the irradiated system could be very bright in the IR. Observations at these wavelengths could be useful in determining the counterpart.

\section{CONCLUSIONS}
\label{sec:conclusions}

We have constructed a model describing the heating effect of a ULX on a Roche lobe filling companion star. We plan to apply this model to the problem of positively identifying ULX optical counterparts, and then use observations of that counterpart to constrain the BH mass and the nature of the companion. 

Our model uses a radiative transfer formulation to account for the X-ray nature of the incident radiation and the distribution of the re-radiated emission. We incorporate the distorted Roche lobe filling geometry of the star and account for the limb and gravity darkening effects. We also include the additional luminosity from an irradiated accretion disk, for two different disk models. 

We have illustrated how our model can be applied in a number of different ways. First, we assume that the donor star is filling its Roche lobe, as suggested by the high X-ray luminosity. We have then shown that for a given spectral type, the effects of irradiation decrease as the mass of the black hole in the system (and hence binary separation) is increased. Measurement of the amplitude and period of the lightcurve of a counterpart will allow us to determine many of the free parameters of the system -- we have shown how our model incorporates both the ellipsoidal and irradiative effects and that we can separate out these components. We have discussed how the accretion disk can be the dominant component in some systems and have shown that measurement of the periodicity of the binary can distinguish between these regimes. We have shown how the importance of the disk is linked with the luminosity class of the companion star, and predict that observations at infrared wavelengths makes it easier to distinguish between different systems without the need for temporal observations.

Finally, we have applied our model to a set of observations of potential candidates for the counterpart of ULX X-7 in NGC 4559. While the available data is not sufficient to draw any firm conclusions, we see that if the observations contain any appreciable accretion disk component then only three of the possible candidates fit our model. These candidates imply the donor star is an early-type MS star. Supergiant companions are only possible for a high inclination and a BH mass of $\sim 1000$\Msun.

\section*{ACKNOWLEDGEMENTS}

We would like to thank the referee (G. Dubus) for his comments, which led to a number of important improvements to this paper. We would also like to thank Richard Mushotzky for interesting discussions.

\appendix

\section{The radiative transfer formulation}
\label{sec:ppmodel}

We consider the effects of radiative transport in the irradiated surface
under X-ray illumination. We consider a plane-parallel model and adopt the
radiative transport formulation of \citet{Milne26} and \citet{Wu01}. Milne's
original formulation was for incident radiation at optical wavelengths and
cooling via emission of optical radiation. This was modified by \citet{Wu01} to
account for incident radiation at X-ray wavelengths. The modification is
important because soft X-rays will be absorbed close to the surface of the star
by neutral and weakly ionized matter via bound-free transitions. Hard X-rays
will only be attenuated at great depths when the matter density is
significantly higher. The soft and hard X-ray components will subsequently have
higher or larger opacities than for the optical radiation. The formulation is
linear and therefore the principle of superposition is applicable. This allows
us to derive the total emission using the irradiated and non-irradiated
components.

\begin{figure}
\centering
\includegraphics{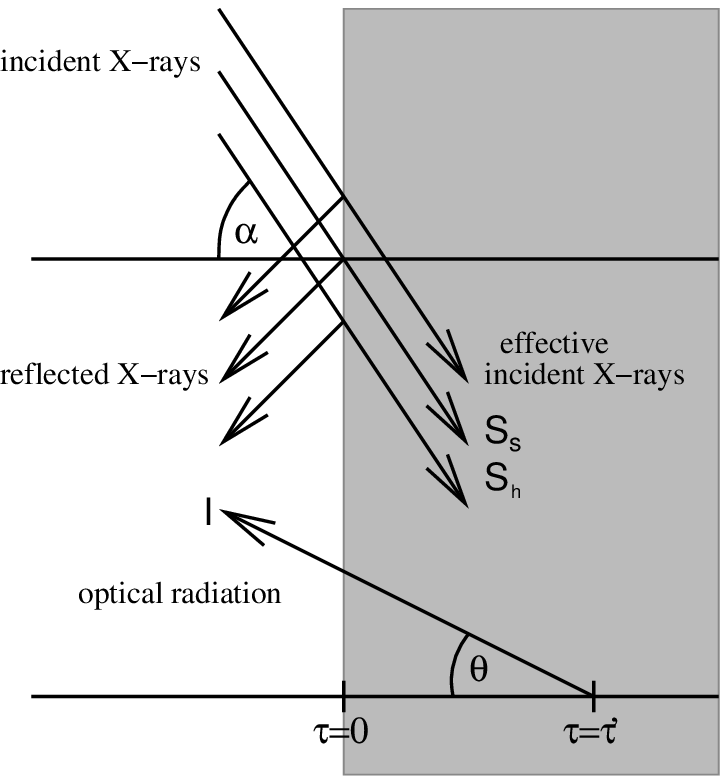}
\caption{The geometry of the plane-parallel model (Wu et al., (2001))}
\label{fig:pp_geom}
\end{figure}

We take the incident radiation to be parallel beams of soft and hard X-rays,
with effective fluxes $\pi S_s$ and $\pi S_h$ per unit area normal to the
beams, and making an angle $\alpha$ to the normal to the stellar surface. The
absorption coefficients of the soft and hard X-rays are $k_s\kappa$ and
$k_h\kappa$ respectively, where $\kappa$ is the absorption coefficient of the
optical radiation ($k_s>1$ and $k_h<1$ defines our soft/hard X-ray convention
in this study, following \citealt{Wu01}). 

\begin{figure}
\centering
\includegraphics[width=3in]{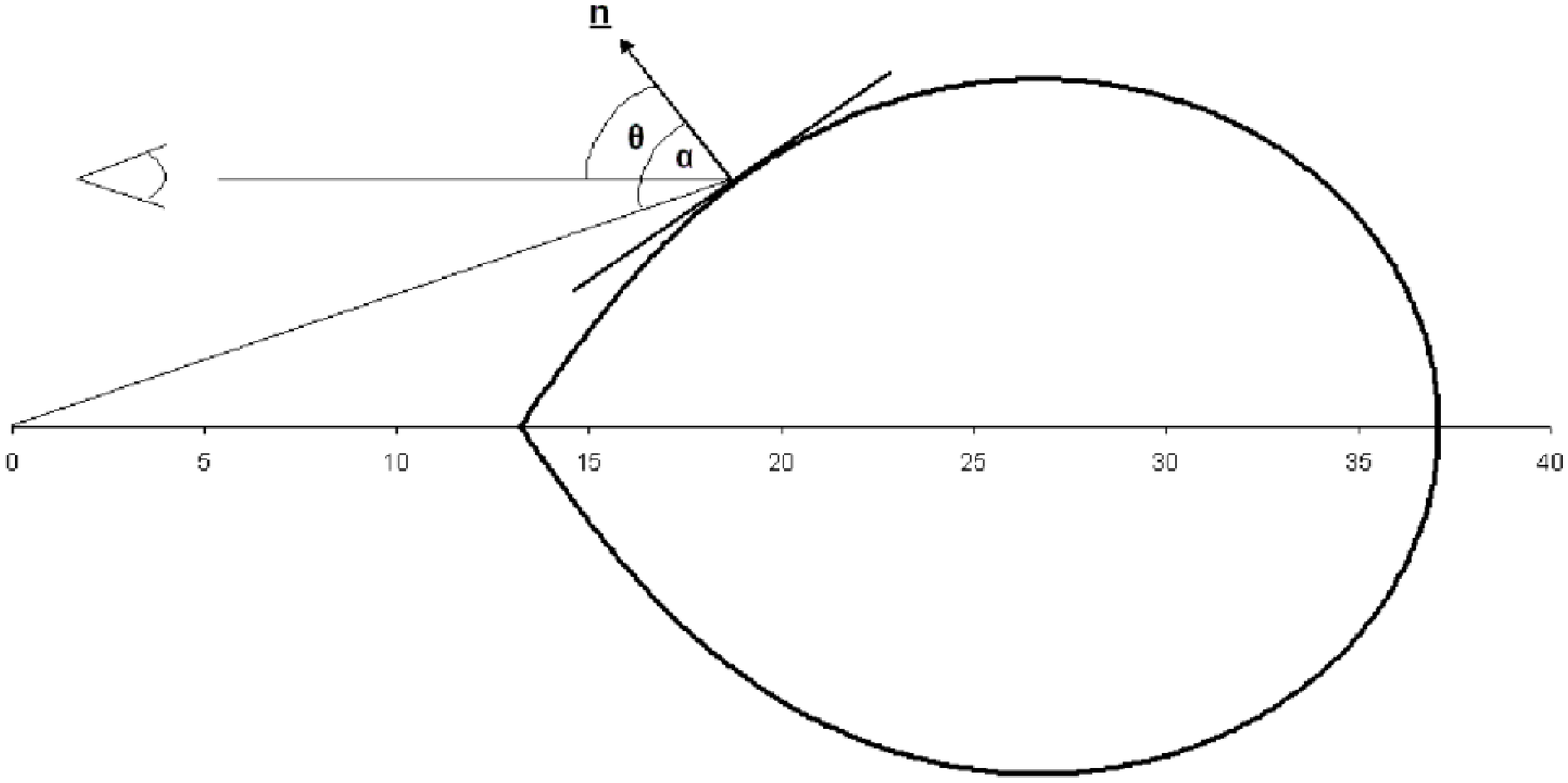}
\caption{The geometry of the binary system}
\label{fig:bin_geom}
\end{figure}

The total blackbody radiation flux is a combination of a component $B_x(\tau)$
as a result of irradiative heating by the incident X-rays and the component of
the radiation from the star in the absence of irradiative heating $B_s(\tau)$,
where $\tau$ is the optical depth. The irradiative heating component
$B_x(\tau)$ was solved in the limit of a semi infinite plane by
the method of successive approximations and was found to be
\begin{equation}
\label{eqn:bx}
B_x(\tau)=a-b_s\exp(-k_s\tau\sec\alpha)-b_h\exp(-k_h\tau\sec\alpha)
\end{equation}
in the second approximation \citep{Wu01}, where $a$, $b_s$ and $b_h$ are
constants to be determined by the boundary conditions. For a semi-infinite slab
opaque at optical wavelengths, the emergent optical/IR radiation in the direction
$\theta$ is the Laplace transform of $B_x(\tau)$
\begin{eqnarray}
I(0,\mu)&=&\lim_{\tau_{tot} \to \infty}\Bigl[\int_0^{\tau{tot}} d\tau B_x(\tau) \exp(-\tau/{\cos\theta})\Bigr] \nonumber \\
&=&a-b_s{\cal A}_s\Bigl[{\cal A}_s+\mu\Bigr]^{-1} - b_h{\cal A}_h\Bigl[{\cal A}_h+\mu\Bigr]^{-1}
\label{eqn:emergrad}
\end{eqnarray}
where ${\cal A}_s$ and ${\cal A}_h$ are $\cos\alpha / k_s$ and $\cos\alpha / k_h$ respectively, and $\mu = \cos\theta$.

Here $a$,$b_s$ and $b_h$ are obtained by solving the radiative-equilibrium and
radiative transfer equations for the conditions $b_s\to 0$ when $S_s\to 0$ and
$b_h \to 0$ when $S_h\to 0$:
\begin{equation}
a={1\over2}\Bigl[k_s S_s {\cal A}_s f_s(\alpha) + k_h S_h {\cal A}_h f_h(\alpha) \Bigr]
\end{equation}
\begin{equation}
b_s={1\over2}k_s S_s \Bigl[{\cal A}_s -{1\over2}\Bigr] f_s(\alpha)
\end{equation}
\begin{equation}
b_h={1\over2}k_h S_h \Bigl[{\cal A}_h -{1\over2}\Bigr] f_h(\alpha)
\end{equation}
where the functions $f_s(\alpha)$ and $f_h(\alpha)$ are given by
\begin{eqnarray}
f_s(\alpha)=\Bigl[1- {\cal A}_s  +  {\cal A}_s   \Bigl({\cal A}_s-{1\over2} \Bigr) \ln(1+k_s\sec{\alpha})\Bigr]^{-1}
\end{eqnarray}
\begin{eqnarray}
f_h(\alpha)=\Bigl[1- {\cal A}_h  +  {\cal A}_h   \Bigl({\cal A}_h-{1\over2} \Bigr) \ln(1+k_h\sec{\alpha})\Bigr]^{-1}
\end{eqnarray}
The hardness of the X-ray source is defined in terms of a hardness parameter $\xi={S_h / S_s}$, with the total X-ray flux $S_x = S_s + S_h$. By expressing $B_x(\tau)$ in terms of this parameter we obtain
\begin{eqnarray}
\label{eqn:bxfinal}
B_x(\tau)={1\over2}S_x \Bigl\{ k_s f_s(\alpha) \Bigl(
{\xi\over{1+\xi}}\Bigr)\Bigl[{\cal A}_s - \Bigl({\cal A}_s-{1\over2}\Bigr)
e^{\tau /{\cal A}_s}\Bigr] \\ \nonumber + k_h f_h(\alpha) \Bigl(
{1\over{1+\xi}}\Bigr)\Bigl[{\cal A}_h - \Bigl({\cal A}_h-{1\over2}\Bigr) e^{\tau
/{\cal A}_h}\Bigr]\Bigr\}\ .
\end{eqnarray}
As the radiative-transfer equations are linear, the local temperature
stratification is given by
\begin{equation}
T(\tau)=\Bigl\{{\pi \over \sigma}[B_x(\tau)+B_s(\tau)]\Bigr\}^{1/4} \equiv
\Bigl({\pi \over \sigma}B(\tau)\Bigr)^{1/4}\ .
\label{eqn:btau}
\end{equation}
The surface temperature of a star is effectively the temperature at an optical
depth of $\tau = 2/3$. Hence, we find when it is viewed at a given inclination angle $\alpha$, the effective temperature of the surface under irradiation is
\begin{equation}
T_{eff}=\Bigl\{{\pi \over \sigma}B_x(2/3)+ T_{unirr}^4\Bigr\}^{1/4}
\end{equation}
where $T_{unirr}$ is the effective temperature in the absence of any
irradiation.


\begin{thebibliography}{}
\bibitem[Allen, 1973]{Allen73} Allen C.W., 1973, {\it Astrophysical Quantities}, (London, UK: Athlone Press)
\bibitem[Bahcall \& Bahcall, 1972]{Bahcall72} Bahcall J.N. \& Bahcall N.A., 1972, ApJ, 178, L1
\bibitem[Begelman, 2002]{Bege02} Begelman M.C., 2002, ApJ, 568, L97
\bibitem[Colbert \& Mushotzky, 1999]{ColMus99} Colbert E.J.M. \& Mushotzky R.F., 1999, ApJ, 519, 89
\bibitem[Cropper et al., 2004]{Cr04} Cropper M., Soria R., Mushotzky R.F., Wu K., Markwardt C.B., Pakull M., 2004, MNRAS, 349, 39
\bibitem[de Jong et al., 1996]{deJong96} de Jong J.A., van Paradijs J., Augusteijn T., 1996, A\&A, 314, 484
\bibitem[Dubus et al., 1999]{Dubus99} Dubus G., Lasota J-P., Hameury J-M., Charles P., MNRAS, 303, 139
\bibitem[Ebisawa et al., 2003]{Ebisawa03} Ebisawa K., \.Zycki P., Kubota A., Mizuno T., Watari K., 2003, ApJ, 597, 780
\bibitem[Fabbiano \& White, 2003]{Fab03} Fabbiano G., White N.E., 2003 in {\it Compact Stellar X-ray Sources}, eds., W. Lewin, W., van der Klis, M., (Cambridge, UK: Cambridge Univ. Press) (astro-ph/0307077)
\bibitem[Fabbiano, 2004]{Fab04} Fabbiano G., 2004, RevMexAA (Serie de Conferencias), 20, 46
\bibitem[Fabrika, 2004]{Fabrika04} Fabrika S. N., 2004, Astrophys. Space Phys. Rev., 12, 1
\bibitem[Gerend \& Boynton, 1976]{Gerend76} Gerend D. \& Boynton P.E., 1976, ApJ, 209, 562
\bibitem[Howarth \& Wilson, 1983]{Howarth83} Howarth I.D. \& Wilson B., 1983, MNRAS, 202, 347
\bibitem[Kaaret et al., 2001]{Kareet01} Kaaret P. et al., 2001, MNRAS, 321, L29
\bibitem[K\"ording et al., 2002]{Kord02} K\"ording E., Falcke H., Markoff S., 2002, A\&A, 382, L13
\bibitem[King et al., 2001]{King01} King A.R., Davies M.B., Ward M.J., Fabbiano G., Elvis M., 2001, ApJ, 552, L109
\bibitem[Lejeune \& Schaerer, 2001]{Lejeune01} Lejeune T. \& Schaerer D., 2001, A\&A, 366, 538
\bibitem[Makishima et al., 2000]{Maki00} Makishima K. et al., 2000, ApJ, 551, L27
\bibitem[Matsumoto et al., 2001]{Matsu01} Matsumoto H. et al., 2001, ApJ, 547, L25
\bibitem[Miller et al., 2003]{Miller03} Miller J.M., Fabbiano G., Miller M.C., Fabian A.C., 2003, ApJ (letters), 585, 37
\bibitem[Milne, 1926]{Milne26} Milne E.A., 1926, MNRAS, 87, 43
\bibitem[Paczy\'nski, 1977]{Paczynski77} Paczy\'nski B., 1977, ApJ, 216, 826
\bibitem[Pakull \& Mirioni, 2002]{Pakull02} Pakull, M.W. \& Mirioni L., 2002, (astro-ph/0202488)
\bibitem[Phillips \& Podsiadlowski, 2002]{Phillips02} Phillips S.N. \& Podsiadlowski P., 2002, MNRAS, 337, 431
\bibitem[Podsiadlowski, 1991]{Podsi91} Podsiadlowski P., 1991, Nat, 350, 136
\bibitem[Ruderman et al., 1989]{Ruder89} Ruderman M., Shamam J., Tavani M., 1989, ApJ, 336, 507
\bibitem[Shakura \& Sunyaev, 1973]{Shak73} Shakura N.I. \& Sunyaev R.A., 1973, A\&A, 24, 337
\bibitem[Soria et al., 2005]{Soria05} Soria R., Cropper M., Pakull M., Mushotzky R., Wu K., 2005, MNRAS, 356, 12
\bibitem[Strohmayer \& Mushotzky, 2003]{Stroh03} Strohmayer T.E. \& Mushotzky R.F., 2003, ApJ, 586, L61
\bibitem[Swartz et al., 2004]{Swartz04} Swartz D.A., Ghosh K.K., Tennant A.F., Wu K., 2004, ApJS, 154, 519
\bibitem[Von Zeipel, 1924]{vZeipel24} von Zeipel H., 1924, MNRAS, 84, 702
\bibitem[Vrtilek et al., 2001]{Vrtilek01} Vrtilek S.D., Quantrell H., Boroson B., Still M., Fiedler H., O'Brien K., McCray R., 2001, ApJ, 549, 522
\bibitem[Wu et al., 2001]{Wu01} Wu K., Soria R., Hunstead R.W., Johnston H.M., 2001, MNRAS, 320, 177

\end{thebibliography}
\end{document}